\documentclass[twocolumn,aps,prb,showpacs,preprintnumbers,amsmath,amssymb, superscriptaddress]{revtex4}

\usepackage{graphicx}
\usepackage{dcolumn}
\usepackage{bm}

\begin{document}


\title{The Thermodynamic Stability of Two Dimensional Crystals with an Extended Coupling Scheme}


\author{D. J. Priour, Jr}
\affiliation{Department of Physics, University of Missouri, Kansas City, Missouri 64110, USA}


\date{\today}

\begin{abstract}
We calculate mean square deviations for crystals in one and two dimensions. 
For the two dimensional lattices, we 
consider several distinct geometries (i.e. square, triangular, and honeycomb), 
and we find the same essential phenomena for 
each lattice structure. We  
investigate the stability of long-range crystalline order for a variety of coupling 
schemes, including short-range exponentially decaying inter-atomic potentials and 
long-range interactions with a power law dependence $r^{-\alpha}$.  For the latter in the 1D case,
we find a critical value $\alpha_{c}^{\mathrm{1D}} = 1.615 \pm 0.005$ for the power law decay exponent 
below which crystalline order is intact, and above which thermal fluctuations 
destroy long-range order when $T > 0$.  The corresponding critical value for two dimensional lattices with 
displacements confined to the plane is $\alpha_{c}^{\mathrm{2D}} = 3.15 \pm 0.025$.  If motion perpendicular to 
the crystal plane is permitted, thermally induced distortions  
diverge rapidly (i.e. linearly) in dual layer systems with local stiffness provided by an extended 
coupling scheme, even if the interaction is long ranged, decaying as a power law in the separation  
between lattice sites.
\end{abstract}

\maketitle


\section{Introduction and Theoretical Techniques}

Crystals are regular periodic physical systems with the atomic 
constituents organized in periodic arrays, where the periodicity is a characteristic 
of the crystal in its equilibrium configuration and a manifestation of long-range 
positional order.  However, the effect of thermal fluctuations must be taken into 
consideration at finite temperatures where thermally excited lattice vibrations 
may degrade or destroy crystalline order.  An early theoretical treatment developed by 
Lindemann~\cite{uno} examined the effect of lattice vibrations in a framework which  
neglected the correlations of atomic motions, but which nonetheless provides a reasonable
description (on an order of magnitude basis) for the melting of three dimensional 
solids.  

A salient component of the Lindemann analysis is the Lindemann criterion 
where the melting of a solid is considered to have taken place when mean square deviations 
from equilibrium exceed a tenth of a lattice constant.  X-ray diffraction data, which may provide a
direct measure of long-range order in a crystal lattice (and hence a means to determine 
temperatures where crystalline order is lost) finds reasonable agreement~\cite{unopuntocinco,unopuntosiete} with the Lindemann  
criterion. The accord is manifest in the finding  
that Bragg peaks corresponding to broken translational symmetry vanish 
when mean square fluctuations from equilibrium (also determined from an analysis of 
X-ray diffraction data) are in the vicinity of a tenth of a lattice constant, as 
specified in the Lindemann result.  

A factor of significance for the effectiveness of the Lindemann criterion is the tendency for 
atoms in three dimensional crystals to have a large number of neighbors [e.g. a dozen 
nearest neighbors in the face of face centered cubic (fcc) lattices].  Hence, mean 
field treatments in the spirit of Weiss molecular mean field theory are more likely 
to provide a reasonable theoretical description 
since statistical fluctuations tend to suppressed somewhat by averaging when a 
relatively large number of 
neighbors are present.

On the other hand, it should 
be understood that apart from the number of nearest neighbors, 
dimensionality is a very important parameter which may affect the thermodynamic 
characteristics and integrity of a crystal lattice to a large degree.
Nano-scale engineering often takes place in systems of low dimensionality such 
as carbon-nanotubes where the length may exceed the width by several orders of 
magnitude; nanotubes tend to be regarded as one dimensional systems.  
Graphene sheets, covalently bonded single layer honeycomb lattices of carbon atoms,
may be considered genuine monolayers, and hence possess strongly two    
dimensional character.  The thermodynamic stability of a system is strongly 
dependent on its dimensionality with statistical 
fluctuations becoming more important for two dimensional systems, and very  
important for essentially 1D structures such as nanotubes.  

An important theoretical result known as the Mermin-Wagner theorem~\cite{dos,tres} 
predicts that as the bulk limit is approached, thermal fluctuations destroy 
long-range crystalline order in the context of 1D lattices.  However, this result does not preclude the 
stabilization of positional order for $T > 0$ if the interaction between 
atomic members is long-ranged (e.g. decaying as a power law in the separation between
positions in the crystal lattice).  In this work, it is our program to examine 
conditions which preserve long range order in low dimensional systems at finite 
temperatures.

In one dimension, the deleterious effect of thermal fluctuations is 
felt most severely, and ultimately only short-range order exists if the 
interatomic interaction is finite in range.  In three 
dimensions, long-range positional order is intact for finite temperatures 
below the melting temperature $T_{m}$.  Two dimensional solids are often regarded as 
an intermediate case where thermal fluctuations are strong enough to destroy 
long-range order as the size of the system is increased, but only in a very 
gradual manner.  Although crystalline order in 2D systems does not survive
in the thermodynamic limit if the interaction is confined to nearest neighbors or is 
otherwise finite in range, a long-range coupling with power law decay may 
stabilize long-range positional order.  In fact, even for one dimensional 
solids, we find a critical decay exponent $\alpha_{c}^{\mathrm{1D}} = 1.615 \pm 0.005$ below which crystalline
order remains stable for $T > 0$, whereas long-range order is only gradually lost 
in the bulk for power law decays where $\alpha > \alpha_{c}$.
Similarly, the corresponding exponent in 2D is $\alpha_{c}^{\mathrm{2D}} = 3.15 \pm 0.025$,  
where long range crystalline order is preserved for $\alpha < \alpha_{c}^{\mathrm{2D}}$, whereas
thermal fluctuations destroy positional order if $\alpha > \alpha_{c}^{\mathrm{2D}}$.  Within the bounds of 
numerical error, we obtain the 
same value for the threshold exponent $\alpha_{c}^{\mathrm{2D}}$ for distinct 
lattice geometries including square lattices, triangular lattices, and honeycomb lattices.

We examine various types of coupling schemes, including very short-ranged interactions 
where atoms interact with only a few nearest neighbors and 
perhaps also next-nearest neighbors.  We also consider extended schemes where there is a finite 
coupling to all neighbors, but where the interaction is still short-ranged,  
with a rapid decay profile, such as that of an inter-atomic potential  with an 
exponential dependence $V \propto e^{-\gamma r}$  where $\gamma^{-1}$ is the finite 
length scale corresponding to the coupling scheme.  
Finally, we also consider a long-ranged algebraically decaying coupling of the 
form $V(r) = r^{-\alpha}$ where $\alpha$ may assume different values, though for 
the energy per atom to be finite in the bulk limit, the exponent must exceed 
threshold values which depend on dimensionality of the lattice.
For single dimensional systems, one must have $\alpha > \alpha_{L}^{1 \mathrm{D}} = 1$ and 
$\alpha > \alpha_{L}^{2 \mathrm{D}} = 2$ for two dimensional crystals.

We report on a calculation of the atomic root mean square deviation $\delta_{\mathrm{RMS}}$ about 
positions of equilibrium in 1D and 2D crystals. In section I, we discuss theoretical methods 
used to calculate the partition function and hence calculate salient thermodynamic quantities by decoupling the vibrational states 
used to gauge the integrity of long-range crystalline order
such as $\delta_{\mathrm{RMS}}$.  In Section II, we examine one dimensional systems, finding 
positional order to be destroyed except in a long-range coupling scheme, (i.e. a  power law dependence where the 
decay exponent $\alpha$ must lie between $\alpha_{L}^{1 \mathrm{D}} = 1$ and an upper bound exponent 
$\alpha_{c}^{1 \mathrm{D}} = 1.615 \pm 0.005$).  In section III, we perform a similar analysis for two dimensional 
square lattices, where we generalize to an extended scheme, and find a gradual destruction 
of crystalline order with increasing system size 
$L$ for short-ranged couplings.  However, we find that a power law decay profile where 
$\alpha < \alpha_{c}^{\mathrm{2D}} = 3.15 \pm 0.025$ is sufficient to maintain 
positional order at finite temperatures.  In addition to the square lattice, in Section IV we also examine triangular and 
honeycomb lattices, finding the ability of a long-range interaction between atoms to 
preserve crystalline order is not affected by the specific type of lattice geometry under 
consideration, and the threshold exponents in all three cases are identical within the bounds of error 
in the calculations. Finally, in section V, we consider motion transverse to the plane of the crystal 
lattice for locally stiff dual layer systems where even if interaction between particles is taken 
to be long-ranged, 
we find the perpendicular motions rapidly compromise long-range order as the size of the 
system is increased.  

A salient component of our treatment is the explicit accounting for atomic motions in 
discrete systems.  The harmonic approximation, which neglects anharmonic terms in the potential  
set up by geometric effects has been tested directly in the context of Monte Carlo 
simulations and found to be accurate to within one part in $10^{3}$ for the systems  
considered~\cite{cuatro}.

Since our interest is in equilibrium thermodynamic characteristics of the system, we begin with the 
lattice potential 
\begin{align}
V = \frac{1}{2} \sum_{i=1}^{N} \sum_{j=1}^{n_{i}} V(r_{ij}),
\end{align}
where $n_{i}$ is the number of neighbors corresponding to each atom in the system indexed with 
the label $i$, and $N$ is the total number of particles contained in the crystal lattice.  
The factor of $1/2$ in the lattice potential expression is present to compensate for double 
counting of the energy associated with individual ``bonds'' between atomic pairs $i$ and $j$.  

For small deviations from equilibrium positions, a ``harmonic approximation'' is possible, and one 
finds instead
\begin{align}
V = \frac{1}{2} \sum_{i=1}^{N} \sum_{j=1}^{n_{i}} \frac{K_{ij}}{2} (l_{ij} - l_{ij}^{0})^{2},
\end{align}
the first nonzero term of a Taylor expansion of $V(r_{ij})$ where $l_{ij} \approx l_{ij}^{0}$,
and $K_{ij}$ is the second derivative of the potential $V(r_{ij})$ at $r_{ij} = l_{ij}^{0}$.
In the results we report on here, we restrict attention to temperatures below those which would cause 
melting in the bulk (determined by the Lindemann criterion), where the harmonic approximation would tend to fare well.  
A primary issue of interest is whether there is any finite temperature 
range where long-range positional order is intact, and our calculations are in the context of temperatures not of the magnitude that 
would disrupt the bonding topology and create dislocations, but thermal regimes considerably below the 
temperature range which might begin to rupture bonds between neighbors.  
For covalent solids such as two dimensional sheets of graphene and carbon nanotubes where energies stored 
in covalent bonds are far in excess of $k_{\mathrm{B}} T$ at 300K, even room temperature may be considered 
a ``low'' temperature in the sense of being considerably below temperature scales where thermal 
fluctuations would perturb the
local bonding scheme in a significant way.

In one dimension, the bonds are collinear, and the potential will remain quadratic as the energies of all bonds 
between atoms are summed.  However, in two dimensional geometries, restoring forces to oppose displacements from 
equilibrium will be exerted in different directions along distinct bond axes between an interacting 
pair of atoms.  As a consequence, it will be 
necessary to make an additional harmonic approximation in order to obtain a quadratic Hamiltonian and 
subsequently exploit translational invariance for the regular lattices we examine.             

In general, a bond length $l_{ij}$ between atoms $i$ and $j$ will appear as 
\begin{align}
l_{ij} = \sqrt{ \begin{array}{c} (\Delta_{ij}^{0x} + \delta_{i}^{x} - \delta_{j}^{x} )^{2} + 
( \Delta_{ij}^{0y} + \delta_{i}^{y} - \delta_{j}^{y})^{2} \\ + 
(\Delta_{ij}^{0z} + \delta_{i}^{z} - \delta_{j}^{z})^{2} \end{array}}
\end{align}
where $\Delta_{ij}^{0x} \equiv (x_{i}^{0} - x_{j}^{0})$, $\Delta_{ij}^{0y} \equiv (y_{i}^{0} - y_{j}^{0})$,
and $\Delta_{ij}^{0z} \equiv (z_{i}^{0} - z_{j}^{0})$.  Thus, the potential energy stored in the bonds 
depends only on the difference of coordinates such as, e.g., $\Delta_{ij}^{0x}$ for the equilibrium $x$ 
coordinate differences and $(\delta_{i}^{x} - \delta_{j}^{x})$ for differences constructed from 
the corresponding shifts from equilibrium.  If the 
latter are sufficiently small in relation to the former, it is appropriate to expand about 
$\Delta_{ij}^{0x}$, $\Delta_{ij}^{0y}$, $\Delta_{ij}^{0z}$, and to quadratic order one will have 
$(l_{ij} - l_{ij}^{0})^{2} \approx \left [ \hat{\Delta}_{ij} \cdot (\vec{\delta}_{i} - \vec{\delta}_{j} ) \right] ^{2}$ where
$\hat{\Delta}_{ij}$ is a unit vector formed by subtracting the position vectors $\mathbf{x}^{0}_{i}$ and 
$\mathbf{x}^{0}_{j}$
corresponding to the atom $i$ and its neighbor $j$, such that 
$\hat{\Delta}_{ij} = (\mathbf{x}_{i}^{0} - \mathbf{x}_{j}^{0})/ \| \mathbf{x}_{i}^{0} - \mathbf{x}_{j}^{0} \|$.
Hence, the atomic potential may be written to quadratic order in $\delta_{i}$ and $\delta_{j}$, and one has in 
particular
\begin{align}
V_{\mathrm{Har}} = \frac{1}{2} \sum_{i=1}^{N} \sum_{j=1}^{n_{i}} \frac{K_{ij}}{2} \left[ \hat{\Delta_{ij}} 
\cdot (\vec{\delta}_{i} - \vec{\delta}_{j} ) \right]^{2}
\end{align}
It will be necessary to solve an eigenvalue problem to decouple the vibrational 
modes.  However, with Fourier analysis, the problem may be reduced to the task of diagonalizing a 
$2 \times 2$ matrix, a $4 \times 4$ matrix in the case of 
a lattice with a honeycomb geometry, or
at most a $6 \times 6$ matrix for the case of the locally stiff dual-layer system, 
even in cases where the coupling scheme is extended to encompass many neighbors for each atomic 
member.  We will consider systems in one dimension where we show that long-range crystalline order may be stabilized 
in the case of slowly decaying power law potentials, but not for localized exponentially decaying coupling schemes.  
We also examine two dimensional lattice geometries, and find similar phenomena; again, a long-range power law decay is needed 
to preserve crystalline order at finite temperatures.

Finally, we are careful to restrict motion to collinear displacements in the context of 1D systems 
and intraplanar motion for the two dimensional crystals.  The lattices are very easily disturbed by 
transverse displacements, and we find that relaxing the collinear and coplanar restrictions in 
1D and 2D yields mean square fluctuations which diverge rapidly with increasing system size 
$N$.  As we find with explicit calculation in section V, this rapid (i.e. at a linear) growth in $N$ occurs even if 
the interaction between atoms is a long-ranged power law decay in locally stiff dual layer crystal geometries.  

We use the results for the eigenvalues for the vibrational modes to calculate thermodynamic properties related to 
crystalline order such as the thermally averaged mean square fluctuations about equilibrium per site, $\delta_{\mathrm{RMS}}$.
As noted elsewhere~\cite{cuatro} and summarized here, the mean square displacements about equilibrium may be calculated in terms of 
the eigenvalues for the vibrational states.  

In terms of the vibrational modes, the lattice energy may be written as 
\begin{align} 
{\mathcal H}^{\mathrm{Har}} = \frac{1}{2} \sum_{\alpha = 1}^{3M} \lambda_{\alpha} c_{\alpha}^{2}
\end{align}
with $M$ the total number of particles contained in the lattice.
The connection between the vibrational states and the mean square fluctuations is 
\begin{align}
\langle \delta_{\mathrm{RMS}} \rangle^{2} = \frac{1}{M} \sum_{i = 1}^{M} \langle (\delta_{i}^{x})^{2} + (\delta_{i}^{y})^{2} + (\delta_{i}^{z})^{2} \rangle
\end{align}
With the eigenvectors indexed with the label $\alpha$ and using, e.g., $\delta_{i}^{x} = \sum_{\alpha = 1}^{3M} c_{\alpha} v_{\alpha}^{ix}$, 
to express the displacements in terms of the eigenvectors, one finds for a specific system configuration 
\begin{align}
\delta^{2} = \frac{1}{M} \sum_{i=1}^{M} \sum_{\alpha=1}^{3M} \sum_{\alpha^{'} = 1}^{3M} \left[ c_{\alpha} c_{\alpha^{'}} \left( v_{\alpha}^{ix} 
v_{\alpha^{'}}^{ix} + v_{\alpha}^{iy} v_{\alpha^{'}}^{iy} + v_{\alpha}^{iz} v_{\alpha^{'}}^{iz} \right ) \right ]
\end{align}
In the thermal average, the factor $c_{\alpha} c_{\alpha^{'}}$ will be as often positive as negative, and hence only in the case $\alpha = \alpha^{'}$ 
will there be a net contribution to the thermal average $\langle \delta_{\mathrm{RMS}} \rangle^{2}$.  If the vectors are taken to be 
normalized, one finds 
\begin{align}
\langle \delta_{\mathrm{RMS}} \rangle ^{2} = \frac{1}{M} \sum_{\alpha = 1}^{3M} \langle c_{\alpha}^{2} \rangle
\end{align} 
With the lattice energy expressed in this way, the partition function becomes of a product of Gaussian integrals,
\begin{align}
Z = \prod_{\alpha = 1}^{3M} \int_{-\infty}^{\infty} e^{- \beta \lambda_{\alpha} c_{\alpha}^{2}/2} d c_{\alpha} = 
\prod_{\alpha = 1}^{3M} \left( \frac{2 \pi \tau}{\lambda_{\alpha}} \right)^{1/2},
\end{align} 
where $\tau \equiv k_{\mathrm{B}} T$.  Finally, a thermal derivative of $Z$ leads to 
\begin{align}
\langle \delta_{\mathrm{RMS}} \rangle^{2} = \frac{d}{d \tau} \textrm{Ln} (Z) = \sum_{\alpha = 1}^{3M} \lambda_{\alpha}^{-1} \tau
\end{align}
Hence, for the mean square deviation, we have 
\begin{align}
\langle \delta_{\mathrm{RMS}} \rangle = \tau^{1/2} \sqrt{ \sum_{\alpha = 1}^{3M}} \equiv \tau^{1/2} \delta_{\mathrm{RMS}}^{n}
\end{align}
The term in the radical is not temperature dependent, but is instead determined by characteristics of the 
lattice geometry and the bonding scheme between atomic members.  In this work, we calculate $\delta_{\mathrm{RMS}}^{n}$, 
a mean square RMS deviation normalized with respect to temperature.
Zero eigenvalues are artifacts of the periodic boundary conditions, correspond to global translations of the 
lattice, and are excluded from the sum.

\section{Systems in One Dimension} 
In the 1D systems we consider, only longitudinal displacements are examined; similarly, for the two dimensional 
geometries, lattice vibrations are considered to be confined to the two dimensional plane with no transverse
motions considered.  Periodic boundary conditions are implemented in both the one and two dimensional
cases.
An important characteristic of systems in one dimension is the 
fact that all bonds are collinear, and hence there is no purely geometric source of anharmonic effects.
The lattice potential energy will have the form
\begin{align}
V = \sum_{l=1}^{n} \sum_{m = 1}^{n} K_{m} (\delta_{l} - \delta_{l+m})^{2}
\label{eq:Eq5}
\end{align}
Since only longitudinal motions are considered, the label ``$x$'' that would normally appear as a 
subscript on the ``$\delta$'' symbols is suppressed.  The sum recorded in Eq.~\ref{eq:Eq5} is configured to avoid the    
redundant summation over bonds, and the counting ``1/2'' factor is not required.
To maximize the number of neighbors coupled to any particular atom while avoiding multiple couplings to the 
same atom via the periodicity condition, we set $n = (N-1)/2$ and we always consider an odd 
number of atomic members. 
   
It is convenient to operate in terms of Fourier components, where we have $\delta_{l} = \sum_{k} e^{ikl} \delta_{k}$; on 
substitution, the expression for the lattice energy has the form
\begin{align}
E = 2 \sum_{k_{x}} \left[ \sum_{m = 1}^{n} K_{m} (1 - \cos mk_{x} ) \right] |\delta_{k_{x}}|^{2},
\end{align}  
which has been diagonalized with the use of Fourier components $\delta_{k}$.  The eigenvalues are given by 
$\lambda_{k_{x}} = 2 {\displaystyle \sum_{m=1}^{n}} K_{m} (1 - \cos km )$, and the normalized thermally induced shift $\delta_{\mathrm{RMS}}^{n}$ has 
the form
\begin{align}
\delta_{\mathrm{RMS}}^{n} = \left( \sum_{k_{x}} \lambda_{k_{x}}^{-1} \right)^{1/2}
\end{align} 
We first examine a localized potential which in the 1D context
 would certainly be expected to yield a divergent mean square fluctuation $\delta_{\mathrm{RMS}}^{n}$
with increasing system size $N$.  As a companion result to gain complementary insight, we also 
calculate the density of states 
corresponding to the system.  One merit of obtaining the vibrational density of states is the fact that it may be 
computed in the thermodynamic limit without encountering divergences
with the aid of Monte Carlo sampling.  On the other hand, the divergence or convergence of
the mean square deviations will be signaled by specific signatures in the low eigenvalue regime of the density of 
states without the need for an extrapolation to the thermodynamic limit.

\begin{figure}
\includegraphics[width=.49\textwidth]{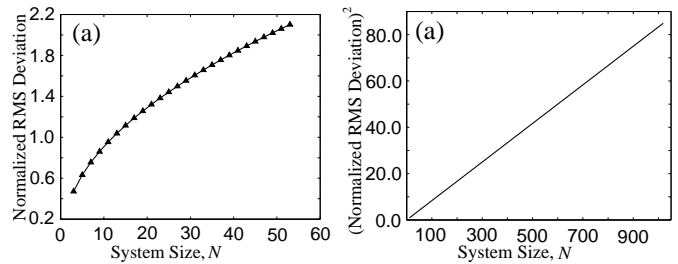}
\caption{\label{fig:Fig1} Mean square deviation curves are graphed
versus system size $N$ in inset (a).  The vertical axis of the graph in inset (b) is $(\delta_{\mathrm{RMS}})^{2}$ to
emphasize the linear behavior of the square of the mean square fluctuations for 
a broader range of system sizes.}
\end{figure}

In the case where interactions are confined to nearest neighbors, the 
eigenvalues $\lambda_{k_{x}}$ have the simple form $2K_{0} (1 - \cos k_{x} x)$.  The corresponding 
thermally averaged $\delta_{\mathrm{RMS}}^{n}$ values in the case of finite systems
are shown in the graph in Fig.~\ref{fig:Fig1}, and there is a steady  rise with 
in the RMS fluctuations with increasing $n$.  
The expansion of the mean square deviations from equilibrium is sub-linear, but it may be shown that the 
increase continues indefinitely (i.e. diverges in the thermodynamic limit) by graphing instead 
$(\delta_{\mathrm{RMS}}^{n})^{2}$, as in the inset of Fig.~\ref{fig:Fig1}.  
The dependence on $N$ quickly reverts to an asymptotically linear increase with 
$N$, and to a good approximation $\delta_{\mathrm{RMS}}^{n} = N^{1/2}$ for moderate to large systems.

We next extend the coupling scheme to many neighbors where the coupling decays at an exponential rate, 
as might be found at least on a qualitative level for a covalently bonded system where the 
rapidly decaying overlap of the orbitals of atomic neighbors (and hence the 
magnitude of the exchange coupling) has an asymptotically exponential 
decay as the separation between the pair of atoms becomes sufficiently large.  The lattice energy will have the form 
\begin{align}
V_{\mathrm{exp}}^{\gamma} = \sum_{l = 1}^{N} \sum_{m = 1}^{n} K_{0} e^{-\gamma m}  ( \delta_{l} - \delta_{l+m})^{2}
\end{align}
Hence in terms of Fourier components, the total lattice potential becomes
\begin{align}
V_{\mathrm{exp}}^{\gamma} =  \sum_{k} \left( 2 K_{0} \sum_{m=1}^{n} e^{-\gamma m} [1 - \cos mk ] \right) 
\end{align}
Again, operating in terms of Fourier components decouples the modes, and the appropriate 
eigenvalues $\lambda_{k}$ are given by
\begin{align}
\lambda_{k} = 2 \sum_{m=1}^{n} e^{-\gamma m} [1 - \cos mk ]
\end{align}
where the prefactor $K_{0}$ has been suppressed.
In addition, the label ``$x$'' on the wave vector has also been suppressed 
for the sake of convenience.
Although the coupling scheme is extended to many neighbors, the potential is in 
an important sense still a local interaction due to its rapid decay, where 
the appropriate length scale is the inverse decay rate $\gamma^{-1}$.
By appealing to the formula for a geometric sum, $r + r^{2} + \ldots + r^{n} = (r-r^{n+1})/(1-r)$ (where $r$ is 
taken to be complex and $\lvert  r \rvert  < 1$) and using the 
fact that $\cos m k = \tfrac{1}{2} (e^{imk} + e^{-imk})$, one may obtain an 
explicit expression for $\lambda_{k}$ which does not require the intermediate summation.
Applying the geometric series formula for a finite series yields     
\begin{align}
&\lambda_{k} = 2 e^{-\gamma} \left( \frac{1 - e^{-\gamma n}}{1 - e^{-\gamma}} \right ) -& \\ \nonumber 
&e^{-\gamma} \left( e^{ik}\left[ \frac{1  - e^{n(-\gamma + ik)}}{1 - e^{-\gamma + ik}} \right] + 
e^{-ik}\left[ \frac{1 - e^{n(-\gamma - ik)}}{1 - e^{-\gamma - ik}} \right] \right)&
\end{align}
Combining the last two fractional terms gives
\begin{align}
\lambda_{k} = 2 \! \left( \! \! \frac{1 - e^{-\gamma n}}{e^{\gamma} - 1} \! \right )\!  - \! \left( \! \frac{ \begin{array}{c} 
\cos k  + e^{-\gamma (n+1)} \cos nk\\ 
 -e^{-\gamma}  - e^{-\gamma n} \cos (n+1)k] \end{array} }{\cosh \gamma - \cos k} \! \! \right ) \! \! , 
\end{align}
a tidier and computationally convenient expression  
to use in calculating the RMS displacements $\delta_{\mathrm{RMS}}^{n}$.  

As $n$ becomes large, terms proportional to $e^{-\gamma n}$ quickly become suppressed by the 
rapid exponential decay.  Hence, for $n \gg \gamma^{-1}$, one will obtain
\begin{align}
\lambda_{k} = \frac{2}{e^{\gamma} - 1} + \frac{e^{-\gamma} - \cos(k)}{\cosh (\gamma) - \cos (k)}
\label{eq:Eq10}
\end{align}

The results are shown in Fig.~\ref{fig:Fig3} for the scaling of $\delta_{\mathrm{RMS}}^{n}$ with respect to the size $N$ 
of the system.  To keep the results for different values of $\gamma$ 
on the same footing, we use the prefactor $K_{0}$ as a normalization of the 
coupling with $K_{0} \equiv e^{\gamma}$.  A similar procedure is also used in calculations involving exponentially 
decaying extended couplings in 2D.  In this manner, the convergence to the results for the case where only 
nearest neighbors are involved in the coupling scheme is easier to see.  

In the main part of the graph, the 
square of the RMS deviation is graphed with respect to system size $N$ for a broad range of system sizes.  
The curves corresponding to the different decay constants are asymptotically linear in the system size 
although the slopes decrease with decreasing $\gamma$ as the coupling becomes longer in range.
The inset of the graph shows a closer view of $\delta_{\mathrm{RMS}}^{n}$.  Each of the curves rises steadily for 
sufficiently large $N$, notwithstanding non-monotonicities for small to moderate $N$ in the case 
$\gamma = 0.25$ where the decay of the interaction is relatively slow.  
In latter case, there is competition between thermal fluctuations and an increase in lattice
rigidity which occurs as the linear crystal grows, providing atoms with more neighbors.  Eventually, however,
$N$ exceeds the length scale $\gamma^{-1}$ of the coupling between atoms, and the balance shifts in 
favor of thermal fluctuations.  The latter increase in importance with increasing $N$ and thus eventually destroy long-range 
crystalline order. 

We also examine the eigenvalue density of states for different decay rates $\gamma$.
With the range of the potential being set by $\gamma^{-1}$, larger values of $\gamma$ would correspond to a more 
rapid decay and a shorter range of the interaction between neighbors.  In calculating the density of states, we use a Monte Carlo sampling process where 
the values of $k$ are not quantized, permitting one to genuinely achieve the bulk limit for the purpose of 
obtaining the density of states.  To obtain a smooth curve a large number of 
eigenvalues (i.e. $2.5 \times 10^{8}$ for the 
histograms corresponding to the exponentially decaying coupling scheme) are sampled. 
The formula given in the continuum limit in Eq.~\ref{eq:Eq10} is the appropriate expression to use for 
$\lambda_{k}$ in the Monte Carlo sampling process.

The normalized density of states for a range of decay constants $\gamma$ appears in Fig.~\ref{fig:Fig3}.   
Panel (a) is a standard plot with the density of states on the vertical axis, while 
to facilitate the viewing of the DOS curves, the logarithm of the DOS curves in shown in panel (b).
Even for relatively long-range cases such as $\gamma = 0.25$, the density of states retains the ``U''-shaped 
profile of the nearest neighbor case.  The latter corresponds effectively to $\gamma = \infty$, and is 
shown in red in the graphs.  For convenience in comparing results, we again choose $K_{0} = e^{\gamma}$ in calculating the 
DOS curves.  The convergence to the $\gamma = \infty$ case with increasing $\gamma$ is 
evident for the case $\gamma = 2.0$, where close agreement with the DOS calculated for the nearest neighbor case 
is evident in panel (b) of Fig.~\ref{fig:Fig1}. In the latter, the ordinate is chosen to be $\log_{10} (\textrm{DOS})$ to help show the 
structure of the density of states curves more clearly. 

\begin{figure}
\includegraphics[width=.49\textwidth]{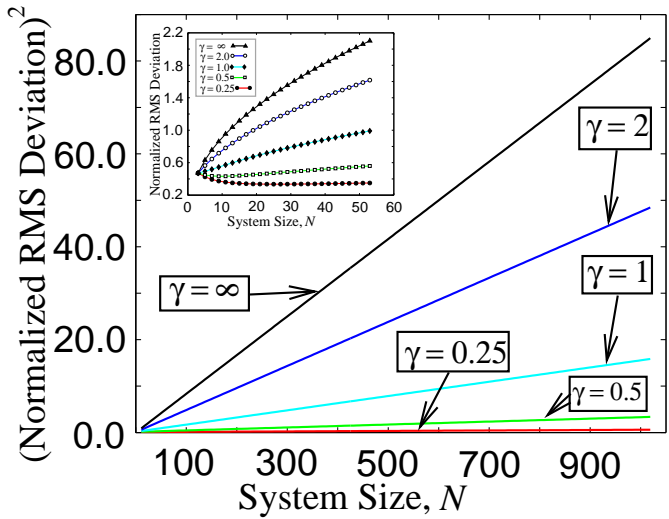}
\caption{\label{fig:Fig2} (Color Online) Mean square deviation curves are shown for a range of 
$\gamma$ values.  The vertical axis of the main graph is $(\delta_{\mathrm{RMS}})^{2}$ to 
emphasize the linear behavior of the square of the mean square fluctuations, while the inset is a   
graph of $\delta_{\mathrm{RMS}}^{n}$ with respect to $N$ for a smaller range of system sizes.
$2.5 \times 10^{8}$ eigenvalues were sampled to calculate the DOS curves.}
\end{figure}

\begin{figure}
\includegraphics[width=.49\textwidth]{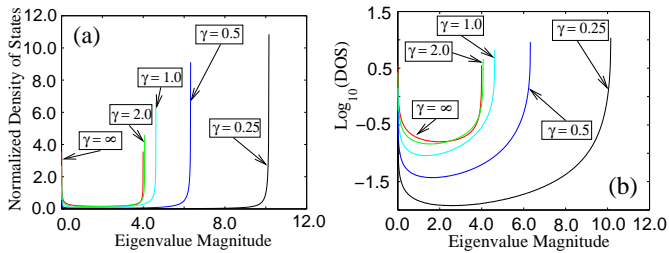}
\caption{\label{fig:Fig3} (Color Online) Eigenvalue Density of States (DOS) for $\gamma$ values
ranging from $\gamma = 0.25$ to $\gamma = 2.0$ to, effectively, $\gamma = \infty$ with the 
DOS on the ordinate for panel (b).
Panel (b) shows the density of states as well, but the 
ordinate is $\log_{10} (\textrm{DOS})$.}
\end{figure}

\begin{figure}
\includegraphics[width=.49\textwidth]{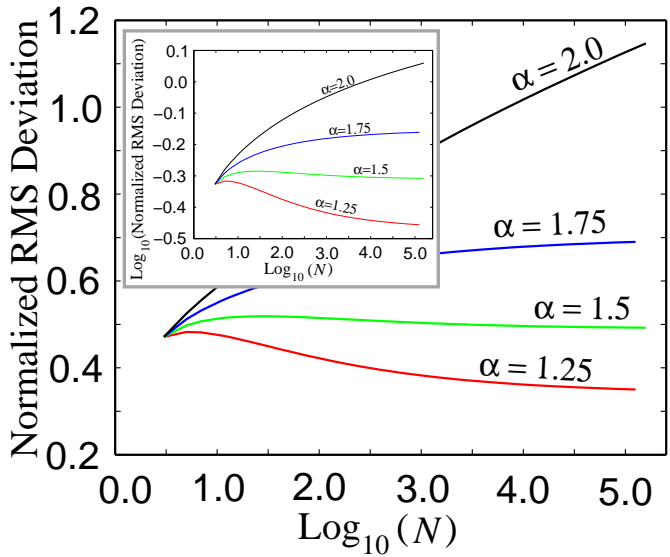}
\caption{\label{fig:Fig4} (Color Online) $\delta_{\mathrm{RMS}}^{n}$ is plotted on a
semi-logarithmic (all logs are base ten)
scale in the main draft for selected values of $\alpha$ for 
one dimensional systems.  The inset is
a log-log graph of the mean square deviations for the same values of $\alpha$ as in the main
graph.}
\end{figure}

Finally, we examine a genuinely long-range inter-atomic potential with a power law decay profile.
The lattice potential energy will have the form 
\begin{align}
E = \sum_{l=1}^{N} \sum_{m=1}^{n} K_{0} m^{-\alpha} (\delta_{l+m} - \delta_{l})^{2}
\end{align}
where $\alpha$ is the decay exponent of the power law interaction ($\alpha > 1$), and 
again $n = (N-1)/2$.
In terms of Fourier components, one will have 
\begin{align}
E = \sum_{k} \sum_{m=1}^{n} 2 K_{0} m^{-\alpha} (1 - \cos mk) \lvert \delta_{k} \rvert^{2}
\label{eq:Eq15}
\end{align}
Hence, the modes are now decoupled with eigenvalues given by $\lambda_{k} = {\displaystyle \sum_{m=1}^{n}} 
m^{-\alpha} (1 - \cos mk)$; we evaluate this expression directly in order to obtain $\delta_{\mathrm{RMS}}^{n}$ 
and the DOS profile appropriate to particular exponent $\alpha$ in the bulk limit.
Again, we calculate $\delta_{\mathrm{RMS}}^{n}$ and generate plots with respect to system 
size.  To test for divergence or convergence in the bulk limit, it is useful also to prepare log-log 
plots (we use base ten logarithms in all cases), and the results appear in the inset of Figure~\ref{fig:Fig4}.  
We examine systems ranging in size from $N = 3$ to on the order of a few hundred thousand atomic members.
A crucial question is whether there is a threshold value $\alpha_{c}$ above $\alpha = 1$ (where the 
lattice energy may diverge with increasing system size) below which long range crystalline 
order is stable with respect to thermal fluctuations in one dimensional lattices.  

To identify $\alpha_{c}^{\mathrm{1D}}$, we calculate the normalized mean square fluctuations $\delta_{\mathrm{RMS}}^{n}$ with respect to 
system size $n$, producing log-log graphs.  The highest value of $\alpha$ where the mean square 
deviations converge is identified as $\alpha_{c}^{\mathrm{1D}}$, the upper limit for the decay exponent in the extended 
power law decay scheme where long-range crystalline order is still supported at finite temperatures.

\begin{figure}
\includegraphics[width=.4\textwidth]{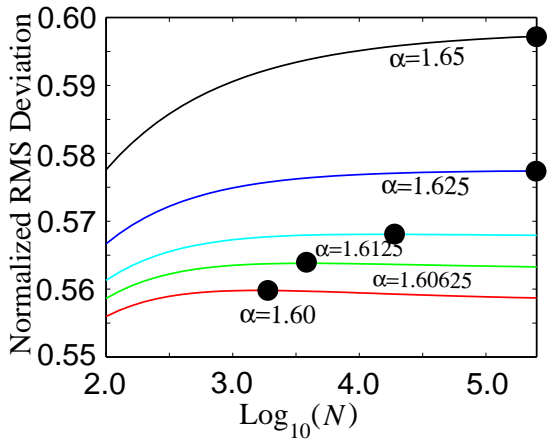}
\caption{\label{fig:Fig5} (Color Online) $\delta_{\mathrm{RMS}}^{n}$ is plotted on a
semi-logarithmic scale for a relatively tight range of $\alpha$ values.  The large dark
circles indicate maxima in the mean square deviation curves; it is concluded that
$\alpha_{c} = 1.615 \pm 0.005$.}
\end{figure}

The mean square                                                                                          
deviations, useful thermodynamic quantities with which to diagnose the presence or absence of  
long-range crystalline order, are shown in Fig.~\ref{fig:Fig4} and Fig.~\ref{fig:Fig5}
(with the abscissa shown as a base ten logarithm over five decades of system
sizes $N$).  In Fig.~\ref{fig:Fig4}, $\delta_{\mathrm{RMS}}^{n}$ curves are shown 
for a relatively wide range of $\alpha$ values.  Over the broad range of systems on 
the horizontal axis, five orders of magnitude, the mean square displacements rise 
monotonically for $\alpha = 2.0$ and $\alpha = 1.75$, while $\delta_{\mathrm{RMS}}^{n}$
decreases steadily for $\alpha = 1.5$ and $\alpha = 1.25$.  The curves suggest a 
decay exponent $\alpha_{c}$ in the vicinity of $\alpha = 1.6$ as a boundary between 
crystals where long-range order is unstable at finite temperatures, and one dimensional 
solid where crystalline order is retained for $T > 0$.  
The inset is the corresponding log-log graph of the mean square fluctuations
plotted for the same $\alpha$ values as in the main graph, which is a semi-logarithmic plot.

In Fig.~\ref{fig:Fig5}, RMS deviation curves are shown for a tighter span of power law
decay exponents 
(ranging from $\alpha = 1.60$ to $\alpha = 1.65$) to identify with greater accuracy the 
numerical value of $\alpha_{c}$.  To facilitate the determination of the 
exponent separating crystals with long-range order and those disrupted by thermal 
fluctuations, we place dark circles over the maxima of the 
$\delta_{\mathrm{RMS}}^{n}$ curves.  For $\alpha > 1.625$, the maxima are located at the 
edge of the plot, consistent with a steady increase (and likely  
divergence in the bulk limit) of the RMS curves.
On the other hand, for $\alpha < 1.6125$, the thermally averaged RMS deviations 
are non-monotonic, reaching a  
maximum for finite values of $N$ and then declining, presumably toward a stable 
bulk value.
We identify the boundary as $\alpha_{c} = 1.615 \pm 0.005$.
It should be emphasized that while long-range order is not supported for decay exponents 
in excess of $\alpha_{c}$, the divergence of $\delta_{\mathrm{RMS}}^{n}$ with increasing 
system size is nonetheless quite slow, sublinear in $\log_{10}(N)$, whereas a strictly logarithmically 
diverging mean square deviation would instead rise at a more rapid linear rate.
 
\begin{figure}
\includegraphics[width=.49\textwidth]{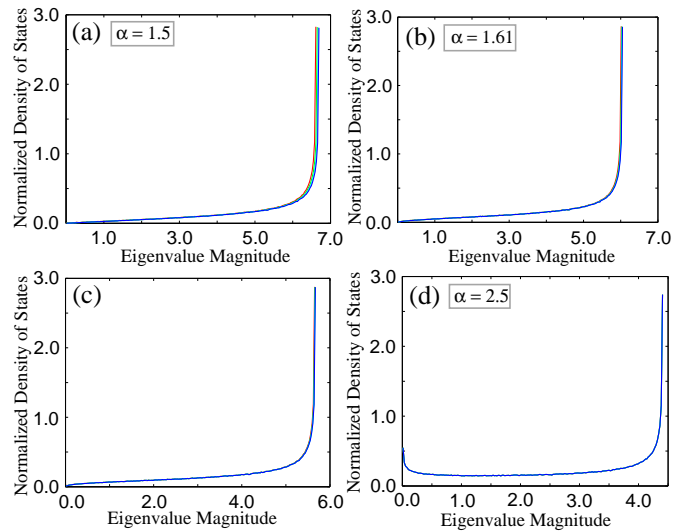}
\caption{\label{fig:Fig6} (Color Online) Eigenvalue Density of States (DOS) for the one 
dimensional lattice with a power law decay where the decay exponent $\alpha$ varies from 
$\alpha = 1.5$ in panel (a) to $\alpha = 2.5$ in panel (d).}
\end{figure}

To obtain information complementary to the RMS fluctuations, we again calculate the eigenvalue 
density of states.  
We also use Monte Carlo sampling where wave numbers are chosen at random,
with uniform probability, to calculate the vibrational density of states with the results shown in Fig.~\ref{fig:Fig6}.
The double sum in Eq.~\ref{eq:Eq15} requires careful consideration, in that one must be certain that enough terms  
have been included in the inner sum that a convergent result is obtained.  To be certain convergence has 
been achieved, 
we prepare eigenvalue histograms for successive doublings of the number of terms contained in the inner sum indexed by 
$m$.  The number of terms which must be included in order to attain suitable convergence 
increases with decreasing $\alpha$ for crystal lattices where the coupling is more slowly
decaying.  In general, however, the oscillatory cosine term in Eq. does act to somewhat hasten 
convergence and hence limit the number of terms which need to be summed.

\section{Two Dimensional Crystals}

\begin{figure}
\includegraphics[width=.49\textwidth]{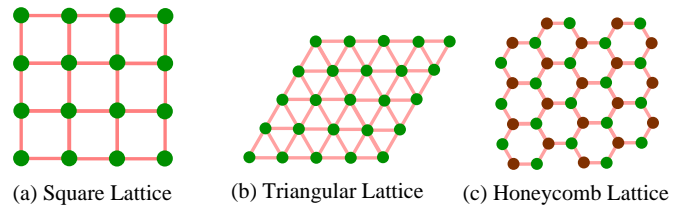}
\caption{\label{fig:Fig7} (Color Online) Three distinct lattice geometries are shown. 
Panel (a) is a portion of a square lattice, panel (b) represents the triangular lattice, and 
panel (c) shows the honeycomb lattice with inequivalent sites represented with distinct colors.}
\end{figure}

For the case of a two dimensional system, the analysis is in many respects parallel to that applied 
for the one dimensional lattices.  However, the additional dimension makes available richer choices 
for the lattice geometry.  We examine various coupling schemes for three types of lattices; the 
square lattice, the triangular lattice, and the honeycomb lattice.  In Fig.~\ref{fig:Fig7}  panel (a) represents 
the square lattice, the triangular lattice is depicted in panel (b), and the honeycomb lattice 
appears in panel (c). A peculiarity of the honeycomb lattice is the presence of inequivalent sites, 
and this characteristic is highlighted in panel (c) of Fig.~\ref{fig:Fig7} where different colors are used in labeling 
the sites.  Although the geometries we examine have different characteristics, the 
essential qualitative characteristics and the most salient physics are found to share much in common.

We first consider the square lattice, and
we initially take into account only interactions between nearest neighbors where at the present level of 
approximation the lattice lacks rigidity.   The lattice energy is 
\begin{align}
E = \frac{K}{2} \sum_{i,j=0}^{N-1} [ (\delta_{i+1}^{x} - \delta_{i}^{x})^{2} + 
(\delta_{ij+1}^{y} - \delta_{ij}^{y})^{2}]
\end{align} 
We express the displacements in terms of Fourier components with, e.g., $\delta_{ij}^{x} 
= {\displaystyle \sum_{\mathbf{k}}} \delta_{\mathbf{k}}^{x} e^{I(k_{x} i + k_{y} j)}$, with $I$ being the 
imaginary unit.  In terms of $\delta_{\mathbf{k}}^{x}$ and $\delta_{\mathbf{k}}^{y}$, the 
energy has the form
\begin{align}
E = \frac{k}{2} \sum_{\mathbf{k}} [ (1 - \cos k_{x} ) \lvert \delta_{\mathbf{k}}^{x} \rvert^{2} + 
(1 - \cos k_{y} ) \lvert \delta_{\mathbf{k}}^{y} \rvert^{2} ];
\label{eq:Eqfloppy}
\end{align}
in this manner the degrees of freedom are decoupled.  Inspection of Eq.~\ref{eq:Eqfloppy} 
reveals that the eigenvalues are $2N$ fold degenerate and identical to the eigenvalues obtained 
for the case of the one dimensional crystal where only interactions between nearest neighbors 
were considered.
Since the eigenvalues are the same as those in the 1D case with interactions only between 
nearest neighbors, crystalline order is readily disrupted by thermal fluctuations.  
Hence, $(\delta_{\mathrm{RMS}}^{n})^{2}$ will scale with $N$ just as was the case for the 1D counterpart.

If one takes into account coupling to next-nearest neighbors as well, then the lattice energy in 
real space is 
\begin{align}
E = \tfrac{K}{2} \! \sum_{i,j=0}^{N-1} \left( \! \! \begin{array}{c} \left[ \hat{x} \cdot 
(\vec{\delta}_{i+1j} - \vec{\delta}_{ij}) \right]^{2} + \left [ \hat{y} \cdot 
( \vec{\delta}_{ij+1} - \vec{\delta}_{ij} ) \right]^{2} \\
+ \left[ \tfrac{1}{\sqrt{2}} ( \hat{x} + \hat{y} ) \cdot \left( \vec{\delta}_{i+1j+1} 
- \vec{\delta}_{ij} \right) \right ]^{2} \\ + \left[ \frac{1}{\sqrt{2}} ( \hat{x} - \hat{y}) 
\cdot \left( \vec{\delta}_{i+1j-1} - \vec{\delta}_{ij} \right) \right]^{2} \end{array} \! \!\right)
\end{align}
Operating in terms of Fourier components, one diagonalizes the matrix
\begin{align}
\left[ \begin{array}{cc} \left( \begin{array}{c} 2 - \cos k_{x} \\ - \cos k_{x} \cos k_{y} \end{array} 
\right) & \sin k_{x} \sin k_{y} \\ 
\sin k_{x} \sin k_{y} & \left( \begin{array}{c} 2 - \cos k_{y} \\ - \cos k_{x} \cos k_{y} \end{array} \right) \end{array} \right]
\end{align}
The eigenvalues are given by 
\begin{align} 
&\lambda_{\pm} = (4 - \cos k_{x} \cos k_{y} - 2 \cos k_{x} \cos k_{y} &  \\
&\pm \sqrt{(\cos k_{x} - \cos k_{y} )^{2} + 4 \sin^{2} k_{x} \sin^{2} k_{y}}&
\end{align}

\begin{figure}
\includegraphics[width=.49\textwidth]{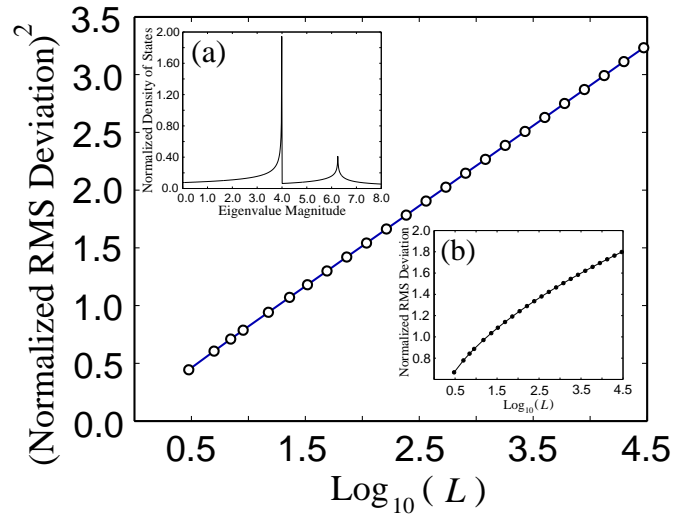}
\caption{\label{fig:Fig8} (Color Online) The mean square displacements 
for the square lattice with nearest and next-nearest neighbor couplings. The main graph shows $(\delta_{\mathrm{RMS}}^{n})^{2}$, 
with the corresponding vibrational density of states graphed in inset (a) and the raw RMS displacements $\delta_{\mathrm{RMS}}^{n}$ displayed 
in inset (b).}
\end{figure}

The results for the mean
square deviations $\delta_{\mathrm{RMS}}^{n}$ appear in Fig.~\ref{fig:Fig8}. 
One may also calculate the vibrational DOS, and the results appear in panel (a) of 
Fig.~\ref{fig:Fig8}.  The introduction of next-nearest neighbor interactions is very effective in 
reducing the deleterious effect of thermal fluctuations on long-range 
crystalline order, though there is still a weak divergence in $\delta_{\mathrm{RMS}}^{n}$ in the
bulk limit.  The square $( \delta_{\mathrm{RMS}}^{n})^{2}$ quickly assumes an asymptotically 
linear form with respect to $\log_{10}N$.
The much slower increase of the RMS deviations with $N$ is
reflected in the DOS profile, where instead of
exhibiting a sharp cusp in the low eigenvalue regime, the 
DOS curve terminates smoothly.  However, the fact that the DOS tends to a finite value as the eigenvalue vanishes is still 
enough to cause a divergence in the mean square displacements from equilibrium.   

We next examine a general case where there are interactions with many neighbors.  
In real space, the energy stored in the lattice has the form
\begin{align}
E = \frac{1}{2} \! \sum_{i,j=-n}^{N-1} \sum_{l,m=-n}^{n} \! \! \! \! \tfrac{1}{2} K(r_{lm}) \! \left( \! \hat{\Delta}_{lm} \! \cdot \!  
[ \vec{\delta}_{i+lj+m} - \vec{\delta}_{ij} ] \! \right )^{2}
\end{align}
where the inner ``$\tfrac{1}{2}$'' factor compensates for multiple counting of bond energies and the choice 
$n = (N-1)/2$ allows each atomic member to interact with all of the atoms contained in   
crystal while avoiding multiple interactions with the same particle.
Since $\vec{\Delta}_{lm} = [l,m]$, the appropriate unit vector directed between particles 
given the labels ``$i+l,j+m$'' and ``$ij$'' is $\vec{\hat{\Delta}}_{lm} = [l,m]/\sqrt{l^{2} + m^{2}}$.
Again, we may decouple the vibrational modes by expressing the coordinate shifts in 
terms of Fourier components.  The lattice potential energy may then be written as
\begin{align}
E =  \frac{1}{2} \sum_{i,j=0}^{N-1} \sum_{l,m=-n}^{n} \tfrac{K(r_{lm})}{2r_{lm}^{2}} \left[ \begin{array}{c} l (\delta_{i+lj+m}^{x} - 
\delta_{ij}^{x} )  + \\ m ( \delta_{i+lj+m}^{y} - \delta_{ij}^{y} ) \end{array}
 \right]^{2} 
\end{align}
with the ``1/2'' factor present to compensate for redundant bond counting.
The range radius $r_{lm}$ is defined 
with $r_{lm} \equiv \sqrt{l^{2} + m^{2}}$, with the full vector given by $\mathbf{r}_{lm} = 
l \hat{x} + m \hat{y}$.
In terms of Fourier components, one will have 
\begin{align}
&E = \frac{1}{2} \sum_{k_{x},k_{y}} \sum_{l,m=-n}^{n} \frac{K(r_{lm})}{r_{lm}^{2}} [1 - \cos (\mathbf{k} \cdot \mathbf{r}_{lm})]&  \\ \nonumber 
&\times \left( \begin{array}{c}  
l^{2} \lvert \delta_{\mathbf{k}}^{x} \rvert^{2}  + m^{2} \lvert \delta_{\mathbf{k}}^{y} \rvert^{2} \\ 
+ ml [\delta_{\mathbf{k}}^{x} \delta_{\mathbf{k}}^{y*} + \delta_{\mathbf{k}}^{x*} \delta_{\mathbf{k}}^{y}] 
\end{array} \right )&
\end{align}
Hence, in order to to decouple the vibrational modes, one must diagonalize the 
matrix
\begin{align}
\tfrac{1}{2} \sum_{l,m=-n}^{n} K(r_{lm}) r_{lm}^{-2} [1 - \cos (\mathbf{k} \cdot \mathbf{r}_{lm})] 
\left[ \begin{array}{cc} l^{2} & ml \\ ml & m^{2} \end{array} \right ],
\end{align}
which may also be written as
\begin{align}
\tfrac{1}{2} \! \! \! \! \! \sum_{l,m=-n}^{n} \! \! \! \! K(r_{lm}) [1 - \cos (\mathbf{k} \! \cdot \! \mathbf{r}_{lm})] \! \! \left [ \! \!
\begin{array}{cc} \hat{\Delta}_{lm}^{x} \hat{\Delta}_{lm}^{x} & \hat{\Delta}_{lm}^{x} \hat{\Delta}_{lm}^{y} \\ 
\hat{\Delta}_{lm}^{y} \hat{\Delta}_{lm}^{x} & \hat{\Delta}_{lm}^{y} \hat{\Delta}_{lm}^{y}
\end{array} \! \right ] \! \! ,
\end{align}
a representation which will prove more compact for more complicated systems such as the honeycomb lattice crystals with  
more than one layer in the direction transverse to the crystal plane, examined in Section V.

\begin{figure}
\includegraphics[width=.49\textwidth]{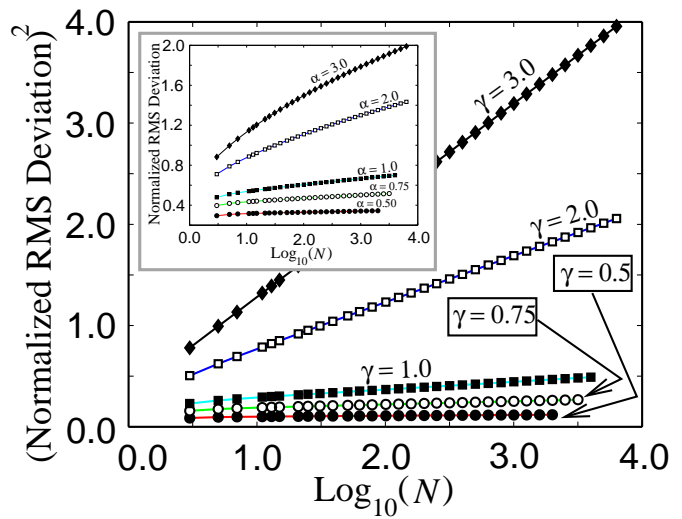}
\caption{\label{fig:Fig9} (Color Online) Mean square deviations for an 
exponentially decaying interaction.  The main graph shows $(\delta_{\mathrm{RMS}}^{n})^{2}$, while
the raw mean square deviations are plotted in the inset of the Figure.} 
\end{figure}

We first turn to the case of an exponentially decaying coupling scheme, and we calculate 
the $\delta_{\mathrm{RMS}}^{n}$ curves with respect to the size $N$ of the system.
The results for the thermally averaged means square displacements 
are shown in Fig.~\ref{fig:Fig9} for a range of different decay constants $\gamma$. 
The computational burden of calculating the auxiliary sum will grow with $N$, but one 
aspect of the exponential decay that is of assistance in the calculations is the fact that the sum may be safely 
truncated when the distance between interacting atoms becomes several times 
greater than the range of the short-ranged coupling (i.e. terms beyond 
beyond $m$ and $l$ pairs such that $\sqrt{l^{2} + m^{2}} \gg \gamma^{-1}$) need not be included.
In particular, we obtain results which are very well converged if we discard terms beyond 20 decay lengths $\gamma^{-1}$.
Ultimately, thermally induced deviations from equilibrium destroy long-range order, and 
the RMS deviations diverge slowly [$(\delta_{\mathrm{RMS}}^{n})^{2}$ again scales linearly 
with $\log_{10}(N)$], but the rate of divergence decreases with decreasing $\gamma$.  In 
particular, as the range $\gamma^{-1}$ of the inter-atomic coupling is increased, 
the slope of the graph of $(\delta_{\mathrm{RMS}}^{n})^{2}$ with respect to 
$\log_{10}(N)$ decreases, although the RMS deviations eventually still diverge in 
the thermodynamic limit.

\begin{figure}
\includegraphics[width=.49\textwidth]{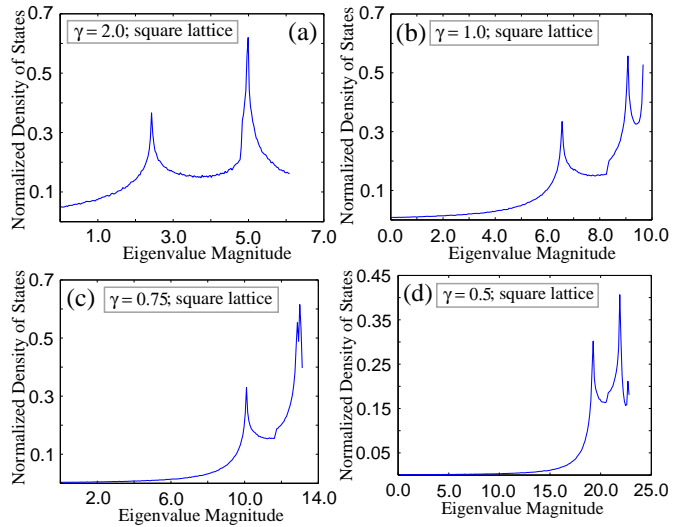}
\caption{\label{fig:Fig10} (Color Online) Vibrational density of states for an 
exponentially decaying coupling.  The DOS curves are plotted for assorted values of the 
decay parameter $\gamma$.}
\end{figure}

The DOS is also calculated, with results appearing in Fig.~\ref{fig:Fig10} for a range of $\gamma$ values.  
We use Monte Carlo sampling to choose $k_{x}$ and $k_{y}$ from a continuum range, and thereby 
operate in the thermodynamic limit for the purpose of calculating the DOS curves.
At least $10^{6}$ eigenvalues are sampled in generating the DOS curves.
The low eigenvalue region of the DOS graph is very similar to the corresponding regime of the 
density of states where only interactions with nearest and next-nearest neighbors are 
included.  

\begin{figure}
\includegraphics[width=.49\textwidth]{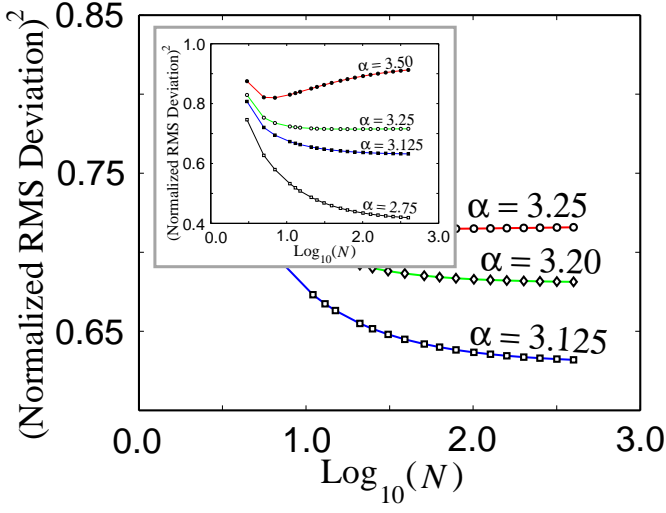}
\caption{\label{fig:Fig11} (Color Online) Mean square deviations for an interaction 
decaying as a power law for decay exponents $\alpha$ near the threshold $\alpha_{c}^{\mathrm{2D}}$.  The 
inset shows a broader view, while the main graph is a closer view of the transition from 
converging to diverging $\delta_{\mathrm{RMS}}^{n}$ curves.}
\end{figure}

Next, we examine the much slower power law decays $K_{ij} = K r_{ij}^{-\alpha}$ in 
the inter-atomic separation $r_{ij}$ where the 
exponent $\alpha$ controls the rate of the decay, long-ranged in 
the sense that there is not a length scale to 
set the range of the coupling.  Again, we first calculate the mean square displacements 
with respect to the size of the system, and then we examine the density of states for 
the eigenvalues.  The $\delta_{\mathrm{RMS}}^{n}$ results are shown in Fig.~\ref{fig:Fig11}.

\begin{figure}
\includegraphics[width=.49\textwidth]{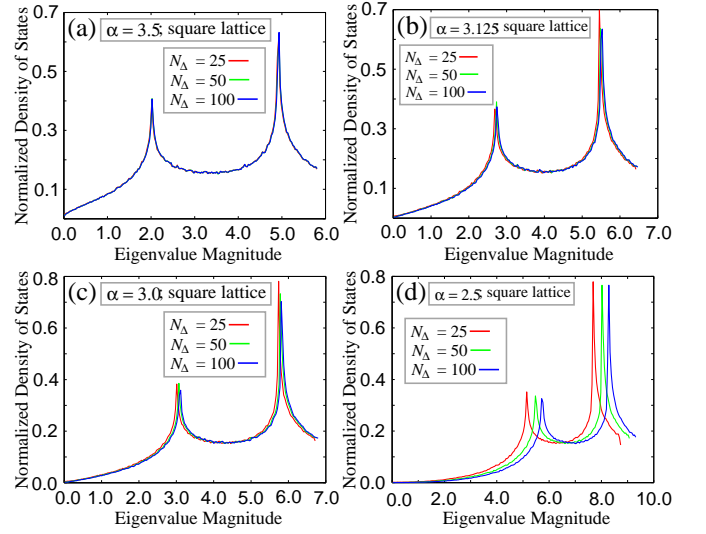}
\caption{\label{fig:Fig12} (Color Online) The vibrational density of states for the two 
dimensional square lattice with a power law decay interaction.  DOS curves for various values of the 
decay parameter $\alpha$ are shown, with traces for different values of the truncation 
length $N_{\Delta}$ on the same plot.}
\end{figure}

Computational subtleties similar to those encountered for the case of the one dimensional 
solid in with a long-ranged coupling scheme must be carefully navigated since the 
higher dimensionality ($d = 2$) will cause the computational burden to grow even more rapidly (nominally as
$L^{4}$ if interactions with all neighbors are included) with system size.  Again, we use Monte Carlo sampling to select wavevectors and accumulate 
eigenvalues to build up the vibrational DOS.  The auxiliary sum 
over dummy indices $l$ and $m$ giving the eigenvalue 
would in principle contain an infinite number of terms (in the bulk limit, each atom in the 
crystal would have an infinite number of neighbors), but we truncate the sum at a 
finite range. The presence of sinusoidal terms in the 
sum, as in the corresponding 1D case, provides an oscillatory element and will hasten the convergence of the sum, thereby reducing the 
computational burden.  We check convergence with respect to the 
truncation range by calculating the DOS with successive doublings of the truncation length $N_{\Delta}$ 
until the DOS profile ceases to change with additional doublings of 
the system size.  The results for the vibrational DOS appear in Fig.~\ref{fig:Fig12}.
One notes that the convergence with respect to the truncation radius is least 
rapid for $\alpha = 2.5$. However, the graphs are relatively well converged for $\alpha = \alpha_{c}^{\mathrm{2D}}$ and 
higher values of the decay exponent.  When $\alpha$ is in the vicinity of $\alpha_{c}^{\mathrm{2D}}$, there is 
very little support in the low eigenvalue regime.  On the other hand, with increasing $\alpha$ the 
interaction decays more rapidly, and ultimately the histogram amplitude in the zero eigenvalue limit rises to a finite value, 
contributing to a divergence in the mean square fluctuations with increasing system size.

\section{Alternate Two Dimensional Geometries}

The treatment in the case of the triangular and honeycomb lattices is very similar to the 
approach used in the case of the square lattice.  
Interestingly, 
the triangular lattice is rigid with only a nearest 
neighbor interaction in the context of the harmonic 
approximation, and the mere inclusion of nearest neighbors is enough to set up quasi-long range 
order where thermally induced fluctuations about equilibrium diverge very slowly [i.e. 
$(\delta_{\mathrm{RMS}}^{n})^{2}$ increases as the logarithm of the system size just as 
in short-ranged extended interactions for the square lattice].
For the triangular lattice, the lattice energy in real space has the form
\begin{align} 
E = \tfrac{K}{2} \sum_{i,j=0}^{N-1} \left( \! \! \begin{array}{c} \left[ \hat{x} \cdot \left (
\vec{\delta}_{i+1j} - \vec{\delta}_{ij} \right ) \right]^{2} + \\ 
\left[ (\tfrac{1}{2} \hat{x} + \tfrac{\sqrt{3}}{2} \hat{y} ) \cdot \left( 
\vec{\delta}_{ij+1} - \vec{\delta}_{ij} \right) \right]^{2} + \\ \left[ (\tfrac{1}{2} \hat{x} - 
\tfrac{\sqrt{3}}{2} \hat{y} ) \cdot \left( \vec{\delta}_{i+1j-1} - \vec{\delta}_{ij} \right )\right]^{2} \end{array} \! \! \right) 
\end{align}
and the eigenvalues for the decoupled vibrational modes are obtained by diagonalizing the $2 \times 2$ matrix
\begin{align}
\left[ \! \! \! \! \begin{array}{cc} \left( \! \! \begin{array}{c} 3  - 2 \cos k_{x} \! - \!  \tfrac{1}{2} \cos k_{y} \\ - 
\tfrac{1}{2} \cos [k_{y} \! - \! k_{x} ] \end{array} \! \! \! \right) & \tfrac{\sqrt{3}}{2} ( \cos [k_{y} \! - \! k_{x} ] - \cos k_{y} ) \\ 
\tfrac{\sqrt{3}}{2} (\cos [k_{y} \! - \! k_{x} ]  \cos k_{y}) & \left( \! \! \begin{array}{c} 3 - \tfrac{3}{2} \cos k_{y} \\ 
- \tfrac{3}{2} \cos [k_{y} \! - \! k_{x} ]  \end{array} \! \! \right) \end{array} \! \! \! \right] \! \! , 
\end{align}
yielding 
\begin{align} 
&\lambda_{\mathbf{k}}^{\pm} = [3 - \cos k_{x} - \cos k_{y} - \cos (k_{y} \! - \! k_{x} ) ]& \\ \nonumber
&\pm \sqrt{\begin{array}{c} \cos^{2} k_{x} + \cos^{2} k_{y} + \cos^{2} (k_{y} \! - \! k_{x}) - \cos k_{x} \cos k_{y} \\
- \cos k_{y} \cos (k_{y} \! - \! k_{x} )  - \cos k_{x} \cos (k_{y} \! - \! k_{x}) \end{array}}& 
\end{align}

\begin{figure}
\includegraphics[width=.49\textwidth]{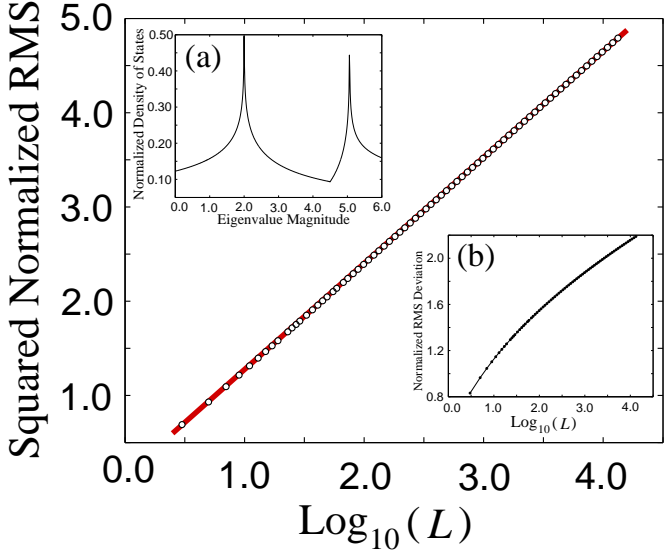}
\caption{\label{fig:Fig13} (Color Online) The mean square deviation for the triangular lattice with nearest neighbor
couplings.  The main graph is a semi-logarithmic plot of $\delta_{\mathrm{RMS}}$, while inset (a) shows the vibrational 
density of states.  A graph of the raw $\delta_{\mathrm{RMS}}^{n}$ appears in panel (b).}
\end{figure}

The results for the mean square fluctuations as well as the vibrational DOS are given in Fig.~\ref{fig:Fig13}. 

\begin{figure}
\includegraphics[width=.45\textwidth]{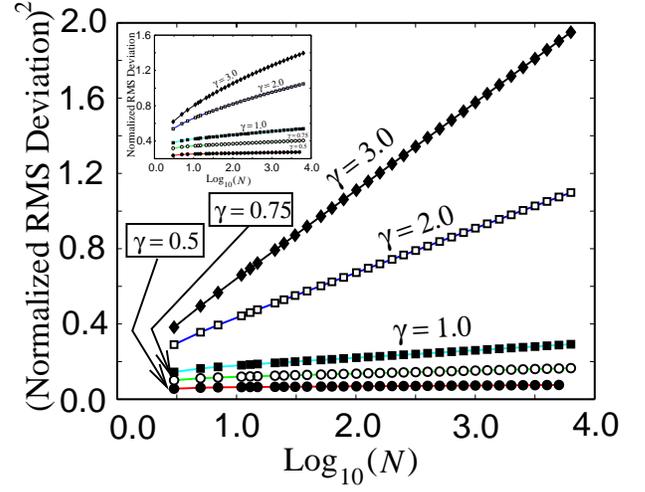}
\caption{\label{fig:Fig14} (Color Online) Normalized Mean square curves and eigenvalue density of 
states for the triangular lattice with an exponential coupling scheme.  The main plot shows $(\delta_{\mathrm{RMS}}^{n})^{2}$ with 
respect to the base ten logarithm of the system size $N$, while inset (a) is the corresponding 
graph for $\delta_{\mathrm{RMS}}^{n}$.  Inset (b) shows the density states curve for the triangular lattice 
for $2.5 \times 10^{8}$ eigenvalues sampled in the bulk limit via Monte Carlo}
\end{figure}

We generalize the nearest-neighbor case to an extended scheme where 
each atomic member may interact with many neighbors.
In real space, the lattice potential energy may be written as 
\begin{align}
E = \frac{1}{2} \sum_{i,j=0}^{N} \sum_{l,m=-n}^{n} \tfrac{1}{2} 
K(r_{lm}) \vec{\hat{\Delta}}_{lm} \! \cdot \! (\vec{\delta}_{i+l,j+m} - \vec{\delta}_{ij}),
\end{align}
where the components of the unit vector $\vec{\hat{\Delta}}_{lm}$ are $\hat{\Delta}_{lm}^{x} = (l+m/2)/r_{lm}$ and 
$\hat{\Delta}_{lm}^{y} = \sqrt{3}m/2r_{lm}$, where $r_{lm} = (l^{2} + m^{2} + lm)^{1/2}$ is the distance separating interacting pairs
in the triangular lattice geometry.
After expressing the displacements in terms of Fourier components, one calculates the eigenvalues of the $2 \times 2$ 
matrix
\begin{align}
\frac{1}{2} \sum_{l,m=-n}^{n} g_{lm} \left[ \! \! \begin{array}{cc} \hat{\Delta}^{x}_{lm} \hat{\Delta}^{x}_{lm} & 
\hat{\Delta}^{x}_{lm} \hat{\Delta}^{y}_{lm} \\ 
\hat{\Delta}^{y}_{lm} \hat{\Delta}^{x}_{lm} & \hat{\Delta}^{y}_{lm} \hat{\Delta}^{y}_{lm} \end{array} \! \! \right ],
\end{align}
where $g_{lm} = [1 - \cos(k_{x}l + k_{y}m)] K(r_{lm})$. 
As in the case of the square lattice, we consider for the triangular lattices an 
exponentially decaying coupling scheme, and the results are shown in Fig.~\ref{fig:Fig14} 
for over four decades of system sizes.
Again, the quantity $(\delta_{\mathrm{RMS}}^{n})^{2}$ increases linearly as $\beta \log_{10} N$ with the slope $\beta$ 
decreasing with decreasing decay rate $\gamma$ (and hence increasing range of the coupling).
 
\begin{figure}
\includegraphics[width=.49\textwidth]{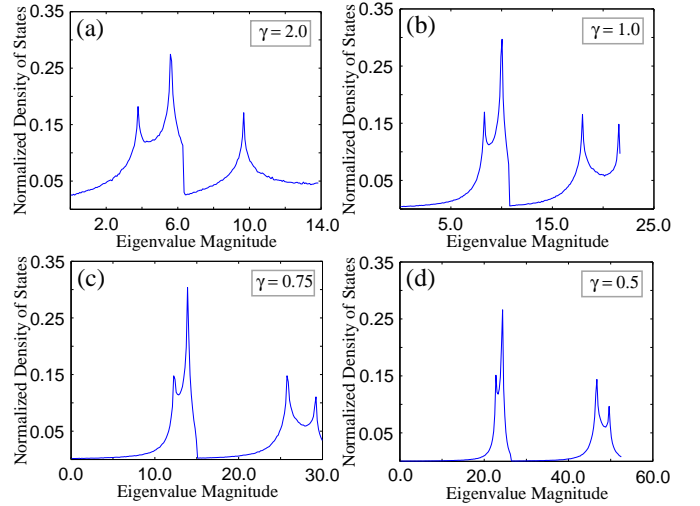}
\caption{\label{fig:Fig15} (Color Online) Density of states curves for $\gamma = 2.0$, 
$\gamma = 1.0$, $\gamma = 0.75$, and $\gamma = 0.50$ in panels (a), (b), (c), and (d) respectively
 for the triangular lattice with an exponentially decaying coupling scheme.}
\end{figure}

We also prepare graphs of the vibrational density of states, shown in the four graphs in Fig.~\ref{fig:Fig15}  
for a range of values of the decay constant $\gamma$.  
The wide separation between the RMS curves corresponding to $\gamma = 2.0$ and $\gamma = 1.0$, $\gamma = 0.75$, and 
$\gamma = 0.5$ is mirrored in the DOS curves where for the smaller decay rates the eigenvalue histogram curve  
intersects the ordinate with very low amplitudes.  On the other hand, for the more rapid decay where $\gamma = 2.0$,
the amplitude in the regime of low eigenvalues is much higher, and the DOS graph resembles that of the nearest neighbor case  
to a much greater degree than DOS profiles corresponding to lower decay rates of the exponential coupling.

\begin{figure}
\includegraphics[width=.49\textwidth]{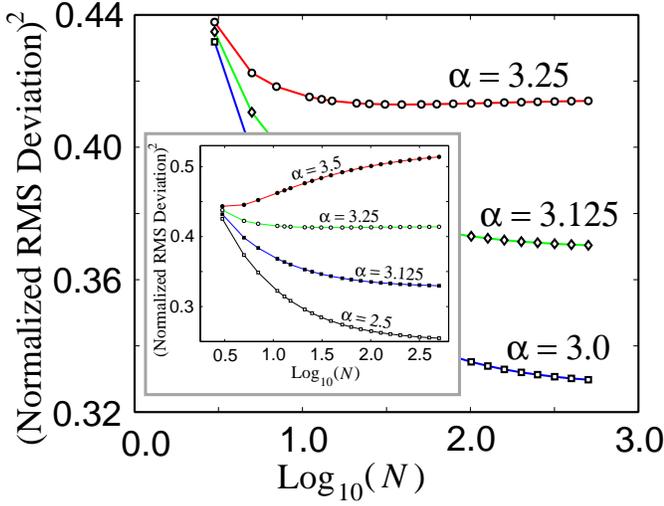}
\caption{\label{fig:Fig16} (Color Online) The square of the normalized RMS deviations for 
various values of the decay exponent $\alpha$, with the inset showing a closer view of the 
$(\delta_{\mathrm{RMS}}^{n})^{2}$ curves.}
\end{figure}

As for the square lattice geometry, we examine a long-ranged power law interaction in the context of 
triangular lattices.  The mean square deviations from equilibrium are graphed in Fig.~\ref{fig:Fig16} with the inset of the plot 
showing a closer view of the $(\delta_{\textrm{RMS}}^{n})^{2}$ curves.  A salient question is if lattices with 
geometries which differ from those of the square lattice will exhibit long-range crystalline order for the 
same range of decay exponents $\alpha$ as in the context of the 
square lattice. We find up to the bounds of error $\alpha_{c}^{\mathrm{2D}} = 3.15 \pm 0.025$ calculated for the triangular lattice to be identical 
to the threshold exponent for the square lattice.

\begin{figure}
\includegraphics[width=.49\textwidth]{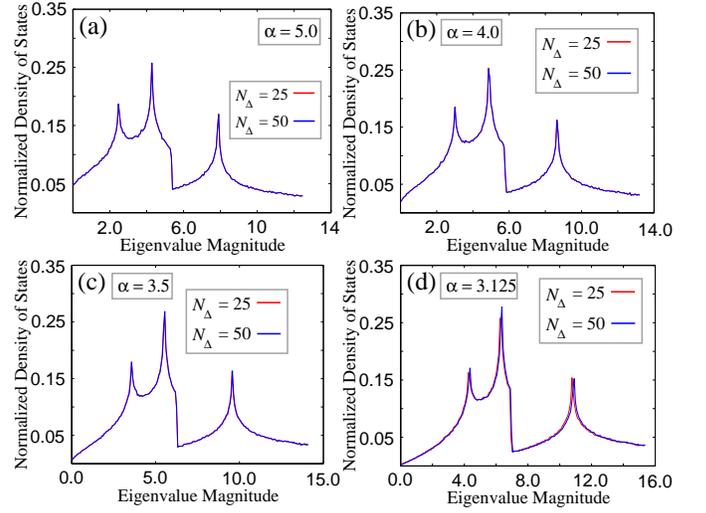}
\caption{\label{fig:Fig17} (Color Online) Eigenvalue histogram curves plotted for $\alpha = 5.0$, 
$\alpha = 4.0$, $\alpha = 3.5$, and $\alpha = 3.125$ respectively. The red trace corresponds to 
a relatively short truncation radius where $N_{\Delta} = 25$, and the truncation is less drastic for 
the blue DOS profiles where $N_{\Delta} = 50$.} 
\end{figure}

We show the corresponding eigenvalue histograms for the power law decay 
for the decay exponents $\alpha = 5.0$, $\alpha = 4.0$, $\alpha = 3.5$, and $\alpha = 3.125$ in panels 
(a), (b), (c), and (d) respectively of Fig.~\ref{fig:Fig17}  While the eigenvalue histograms plotted in the panel (a) and panel (b)
correspond to decay exponents significantly higher than $\alpha_{c}^{\mathrm{2D}}$, the DOS curve in panel (c),     
is plotted for a decay exponent only slightly above the threshold value, and the histogram in panel (d) of Fig. corresponds 
to a value of $\alpha$ just below (though very nearly equal) to $\alpha_{c}^{\mathrm{2D}}$.
Whereas the DOS curves in panel (a) and (b) clearly tend to a finite value as the eigenvalue approaches zero, the 
amplitude for the slower decay $\alpha = 3.125$ and tends to zero in the the limit that the 
eigenvalue is very small; for the case $\alpha = 3.5$, the amplitude reaches a finite
but very small value in the zero eigenvalue limit. 
 A DOS amplitude tending to zero in the low eigenvalue limit, as seen for $\alpha = 3.125$ is 
consistent with the preservation of long-range 
crystalline order indicated in the convergence of the mean square deviations graphed in Fig.~\ref{fig:Fig16}. 

We examine the honeycomb lattice, which differs from the square in triangular lattices in that it possesses 
two inequivalent sites (labeled ``A'' and ``B'' for convenience).
We again appeal to translational invariance, operating in terms of Fourier components, to 
decouple the vibrational modes for the honeycomb lattice.
The relationship of sites of type ``A'' and ``B'' to nearest neighbors is 
illustrated in Fig.~\ref{fig:Fig20}. 
\begin{figure}
\includegraphics[width=.49\textwidth]{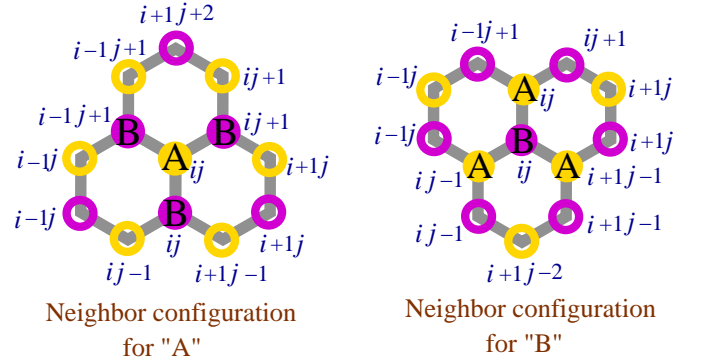}
\caption{\label{fig:Fig18} (Color Online) Labeling and indexing scheme for the honeycomb lattice 
for inequivalent sites labeled ``A'' and ``B'', and their immediate vicinity.}
\end{figure}

Following this labeling convention, the lattice energy in real space has the form 
\begin{align}
E \! = \! \tfrac{K}{2} \! \sum_{i,j=0}^{N-1}  \left \{ \! \! \begin{array}{c} 
\left[ \left( \tfrac{\sqrt{3}}{2}\hat{x} + \tfrac{1}{2} \hat{y} \right) \! \cdot \!  
\left( \vec{\delta}_{ij}^{\mathrm{A}} - \vec{\delta}_{ij+1}^{\mathrm{B}} \right) \right ]^{2} + \\
 \left[ \left( -\tfrac{\sqrt{3}}{2} \hat{x} + \tfrac{1}{2} \hat{y} \right) \! \cdot \! 
\left( \vec{\delta}_{ij}^{\mathrm{A}} - \vec{\delta}_{i-1j+1}^{\mathrm{B}} \right) \right]^{2} 
\\ + \left[ -\hat{y} \! \cdot \! \left ( \vec{\delta}_{ij}^{\mathrm{A}} - 
\vec{\delta}_{ij}^{\mathrm{B}} \right) \right]^{2}
\end{array} \! \! \! \right \}
\end{align}
where it is sufficient to sum over the three bonds surrounding the atoms labeled ``A'' 
with no factor of $\frac{1}{2}$ needed to compensate for double counting.
In Fourier space, the energy stored in the lattice has the form
\begin{align}
E = \tfrac{K}{2} \! \sum_{\mathbf{k}} \! \left \{ \! \! \! \begin{array}{c} 
\tfrac{3}{2} \left( \lvert \delta_{\mathbf{k}}^{\mathrm{A} x} \rvert^{2} + 
\lvert \delta_{\mathbf{k}}^{\mathrm{A} y}
\rvert^{2} + \lvert \delta_{\mathbf{k}}^{\mathrm{B} x} \rvert^{2} + 
\lvert \delta_{\mathbf{k}}^{By} \rvert^{2} \right) \\
-\tfrac{1}{4} ( 1 + e^{-ik_{x}} ) e^{ik_{y}} \! \! \left (3 \delta_{\mathbf{k}}^{\mathrm{B} x} 
\delta_{\mathbf{k}}^{\mathrm{A} x*} + \delta_{\mathbf{k}}^{\mathrm{B} y} \delta_{\mathbf{k}}^{\mathrm{A} x*}
\right) \\
-\tfrac{1}{4} (1 + e^{ik_{x}} ) e^{-ik_{y}} \! \! \left( 3 \delta_{\mathbf{k}}^{\mathrm{B} x*} 
\delta_{\mathbf{k}}^{\mathrm{A} y} + \delta_{\mathbf{k}}^{\mathrm{B} y*} \delta_{\mathbf{k}}^{\mathrm{A} x}
\right)   \\
+ \tfrac{\sqrt{3}}{4} (e^{-ik_{x}} - 1) e^{ik_{y}} \! \! \left( \delta_{\mathbf{k}}^{\mathrm{B} x} 
\delta_{\mathbf{k}}^{\mathrm{A} y*} + \delta_{\mathbf{k}}^{\mathrm{B} y} \delta_{\mathbf{k}}^{\mathbf{A} x*} 
\right) \\ 
+ \tfrac{\sqrt{3}}{4} (e^{ik_{x}} - 1) e^{-ik_{y}} \! \! \left( \delta_{\mathbf{k}}^{\mathrm{B} x*} 
\delta_{\mathbf{k}}^{\mathrm{A} y} + \delta_{\mathbf{k}}^{\mathrm{B} y*}        
\delta_{\mathbf{k}}^{\mathrm{A} x} \right) \\ - \delta_{\mathbf{k}}^{\mathrm{A}y} \delta_{\mathbf{k}}^{\mathrm{B}y*} - 
\delta_{\mathbf{k}}^{\mathrm{A} y *} \delta_{\mathbf{k}}^{\mathrm{B} y} 
\end{array} \! \! \! \! \! \right \}
\end{align}
In addition to Fourier decomposition, the diagonalization of a $4 \times 4$ matrix 
will be necessary to completely decouple the vibrational modes appropriate to the honeycomb lattice 
with the nearest neighbor coupling scheme; the matrix in question is
\begin{align}
\left[ \begin{array}{cccc} c_{\mathrm{A}x \mathrm{A}x} & c_{\mathrm{A}x \mathrm{A}y} & 
c_{\mathrm{A} x \mathrm{B} x} & c_{\mathrm{A} x \mathrm{B} y}  \\
c_{\mathrm{A} y \mathrm{A} x} & c_{\mathrm{A} y \mathrm{A} y} & 
c_{\mathrm{A} y \mathrm{B} x} & c_{\mathrm{A} y \mathrm{B} y} \\
c_{\mathrm{B} x \mathrm{A} x} & c_{\mathrm{B} x \mathrm{A} y} &
c_{\mathrm{B} x \mathrm{B} x} & c_{\mathrm{B} x \mathrm{B} y} \\
c_{\mathrm{B} y \mathrm{A} x} & c_{\mathrm{B} y \mathrm{A} y} & 
c_{\mathrm{B} y \mathrm{B} x} & c_{\mathrm{B} y \mathrm{B} y}
\end{array} \right]  = \left[ \begin{array}{cc} \hat{A} & \hat{B} \\ \hat{B}^{\dagger} & \hat{A} 
\end{array} \right] 
\end{align}
where $\hat{A}$ and $\hat{B}$, and $\hat{B}^{\dagger}$ is the 
Hermitian conjugate of $\hat{B}$.  The  sub-matrices $\hat{A}$ and $\hat{B}$ are given by 
\begin{align}
\hat{A} = \left [ \begin{array}{cc} \tfrac{3}{2} & 0 \\ 0 & \tfrac{3}{2} \end{array} \right ]
\end{align}
and
\begin{align}
\hat{B} = e^{ik_{y}} \! \left[ \! \! \begin{array}{cc} -\tfrac{3}{4} (1 + e^{-ik_{x}} ) & 
\tfrac{\sqrt{3}}{4}  (e^{-ik_{x}} - 1) \\ 
\tfrac{\sqrt{3}}{4} (e^{-ik_{x}} - 1) & -e^{-ik_{y}} - \tfrac{1}{4}(1 + e^{-ik_{x}})  \end{array} \! \! \right]
\end{align}

However, to obtain a crystal which is locally stiff, 
one must examine an extended scheme 
where each atomic member interactions with many neighbors.  In real space, the lattice energy may 
be expressed as 
\begin{align}
&E \! \! =  \! \! \sum_{i,j = 0}^{N}& \\ \nonumber
&\sum_{l,m=-n}^{n} \! \! \left( \! \! \! \begin{array}{c} \tfrac{1}{2} \! K(r_{lm}^{t}) \! \left[ \! 
(\hat{\Delta}_{lm}^{(t)x}  \hat{x} \! + \! \hat{\Delta}_{lm}^{(t)y} \hat{y} )  \! \cdot \!  (  \vec{\delta}_{i+lj+m}^{A} \! - \! 
\vec{\delta}_{ij}^{A} ) \! \right]^{2} + \\ \tfrac{1}{2} K(r_{lm}^{t}) \! \! 
\left[ \! (\hat{\Delta}_{lm}^{(t)x} \hat{x} \! + \! \hat{\Delta}_{lm}^{(t)y} \hat{y} )  \! \cdot \!  (  \vec{\delta}_{i+lj+m}^{B} \! - \! 
\vec{\delta}_{ij}^{B} ) \! \right]^{2} + \\ \!  K(r_{lm}^{\mathrm{ab}}) \! \! \left[ \! (\hat{\Delta}_{lm}^{(\mathrm{ab})x} \hat{x} \! + \! 
\hat{\Delta}_{lm}^{(\mathrm{ab})y} \hat{y}  )  \! \cdot  \! ( \vec{\delta}_{i+lj+m}^{B} \! - \!
\vec{\delta}_{ij}^{A} ) \! \right]^{2} \end{array} \! \! \! \! \right) &
\end{align}
where the first and second terms in the sum take into account interactions between atoms labeled 
``A'' and ``B'', respectively; the identical form of ``A-A'' and ``B-B'' interaction terms is due 
to the fact that the ``A'' and ``B'' species both define triangular lattices, as illustrated in Fig.
We take the lattice constant to be unity, and the components of the unit vector $\vec{\hat{\Delta}}^{t}_{lm}$ appropriate to the 
triangular sublattices are  $\hat{\Delta}_{lm}^{x(t)} = \sqrt{3}(l + m/2)/r_{lm}^{t}$ and $\hat{\Delta}_{lm}^{y(t)} =(3m/2)/r_{lm}^{t}$, where 
$r_{lm}^{t} = (3[l^{2} + m^{2} + lm])^{1/2}$.
On the other hand, the components of $\vec{\hat{\Delta}}^{\mathrm{ab}}_{lm}$ used in calculating interactions between 
``A'' and ``B'' atoms are given by
$\hat{\Delta}_{lm}^{x(\mathrm{ab})} = \sqrt{3}(l + m/2)/r^{\mathrm{ab}}$ and 
$\hat{\Delta}_{lm}^{y (\mathrm{ab})} = (\tfrac{3}{2} m - 1)/r^{\mathrm{ab}}_{lm}$, where 
$r^{\mathrm{ab}}_{lm} = (3l^{2} + 3m^{2} +3lm -3m + 1)^{1/2}$

The exploitation of translational invariance by expressing the displacements in terms of Fourier 
components reduces the decoupling of the vibrational modes to the diagonalization of the $4 \times 4$ matrix $\left[ \begin{array}{cc} 
\hat{A} & \hat{B} \\ \hat{B}^{\dagger} & \hat{A} \end{array} \right]$,
where the sub-matrices are given by 
\begin{align} 
&\hat{A} = \sum_{l,m=-n}^{n}  g_{lm}
 \left[ \! \! \begin{array}{cc} \hat{\Delta}^{x (t)}_{lm}            
\hat{\Delta}^{x (t)}_{lm} & \hat{\Delta}^{x (t)}_{lm} \hat{\Delta}^{y (t)}_{lm} \\ 
\hat{\Delta}^{x (t)}_{lm} \hat{\Delta}^{y (t)}_{lm} & \hat{\Delta}^{y (t)}_{lm} \hat{\Delta}^{y (t)}_{lm} \end{array} \! \! \right ]& 
\\ \nonumber
&+ K(r_{lm}^{\mathrm{ab}}) \left[ \begin{array}{cc} \hat{\Delta}_{lm}^{x(\mathrm{ab})} \hat{\Delta}_{lm}^{x(\mathrm{ab})}
& \hat{\Delta}_{lm}^{x(\mathrm{ab})} \hat{\Delta}_{lm}^{y(\mathrm{ab})} \\
\hat{\Delta}_{lm}^{y(\mathrm{ab})} \hat{\Delta}_{lm}^{x(\mathrm{ab})}
& \hat{\Delta}_{lm}^{y(\mathrm{ab})} \hat{\Delta}_{lm}^{y(\mathrm{ab})} \end{array}\right]&
\end{align}
for $\hat{A}$ and 
\begin{align}
\hat{B} \!  = \! \! \! \! \sum_{l,m=-n}^{n} \! \! e^{I(k_{x}l + k_{y}m)} \! K(r_{lm}^{\mathrm{ab}}) \! \!
\left[ \! \! \! \begin{array}{cc} \hat{\Delta}_{lm}^{x(\mathrm{ab})} \hat{\Delta}_{lm}^{x(\mathrm{ab})} 
& \hat{\Delta}_{lm}^{x(\mathrm{ab})} \hat{\Delta}_{lm}^{y(\mathrm{ab})} \\
\hat{\Delta}_{lm}^{y(\mathrm{ab})} \hat{\Delta}_{lm}^{x(\mathrm{ab})}
& \hat{\Delta}_{lm}^{y(\mathrm{ab})} \hat{\Delta}_{lm}^{y(\mathrm{ab})}\end{array} \! \! \! \right]
\end{align}
where $g_{lm} \equiv (1 - \cos[k_{x} l + k_{y}m]) K(r_{lm}^{t})$, and 
$g_{00} = 0$ to exclude self-interactions in the ``A-A'' and the ``B-B'' coupled pairs.

\begin{figure}
\includegraphics[width=.35\textwidth]{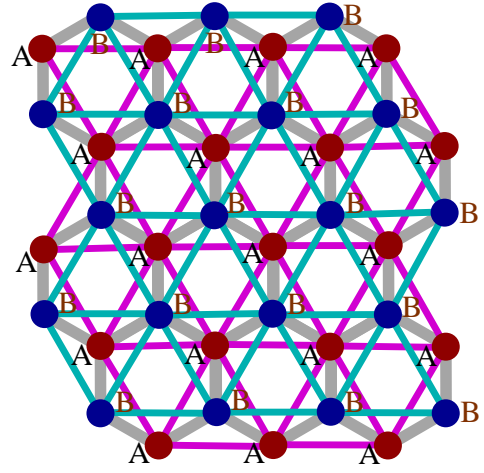}
\caption{\label{fig:Fig19} (Color Online) Honeycomb lattice geometry showing the 
interpenetrating triangular lattices defined by the inequivalent sites labeled 
``A'' and ``B''.}
\end{figure}

As we did for the square and triangular lattices, we examine a short-ranged exponential 
interaction between atoms in the lattice with a length scale given by $\gamma^{-1}$.  
On the other hand, as for the preceding two lattice geometries, we also consider a 
long-ranged power law decay with $K(r) = K_{0} r^{-\alpha}$.  The $\delta_{\mathrm{RMS}}^{n}$
results for the mean square deviations for the exponential decay scheme are shown in Fig.~\ref{fig:Fig20} for a range of $\gamma$ values.
As in the cases of the square and triangular lattices in the extended schemes, increasing the range $\gamma^{-1}$ 
of the coupling slows (but does not halt) the rate of divergence of the mean square fluctuations.  The large separation 
between the RMS curves corresponding to $\gamma = 2.0$ and the slower decays $\gamma = 1.0$, $\gamma = 0.75$, and $\gamma = 0.50$ is  
consistent with changes in the density of states curves where the histogram amplitude in the low eigenvalue regime is   
sharply diminished as $\gamma$ decreases from $\gamma = 2.0$ to $\gamma = 1.0$.

In the case of a power law decay, results for the mean square deviations are shown in the semi-logarithmic graphs in Fig.~\ref{fig:Fig23}
where the systems sizes considered span two decades,
with a closer view for a more restricted range of the decay exponent $\alpha$ in the main graph; the inset is a graph of RMS curves 
for a broader set of $\alpha$ values.  As in the cases of the square and triangular lattices, the threshold exponent 
$\alpha_{c}^{\mathrm{2D}}$ is determined by examining whether the RMS curves converge or diverge for very large system sizes.
In agreement with the square and triangular geometries, we find $\alpha_{c}^{\mathrm{2D}} = 3.15 \pm 0.025$ for the 
critical decay exponent.

The eigenvalue histogram curves shown in Fig.~\ref{fig:Fig23} are consistent with the behavior of the mean square deviation 
curves given in Fig.~\ref{fig:Fig22}.  For the more rapid decays $\alpha = 4.0$ and $\alpha = 4.5$, the amplitude of the 
density of states is finite, which eventually contributes to a divergence in $\delta_{\mathrm{RMS}}^{n}$.  
The divergence in the mean square fluctuations is much slower for $\alpha = 3.5$, a characteristic which is echoed in the 
eigenvalue histogram in panel (c) of Fig.~\ref{fig:Fig23}, where in the zero eigenvalue limit the histogram amplitude is finite 
but very small.  Finally, for $\alpha = 3.125$, just below $\alpha_{c}^{2D}$, the density of states curve tends to zero in the low 
eigenvalue limit.

\begin{figure}
\includegraphics[width=.49\textwidth]{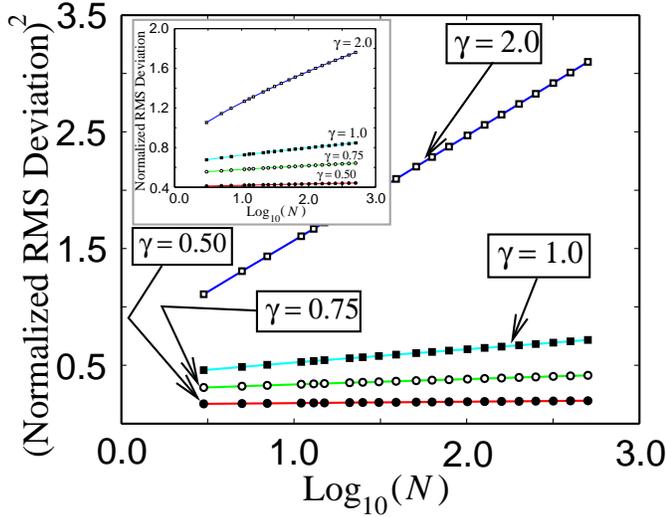}
\caption{\label{fig:Fig20} (Color Online) Mean square deviations for the honeycomb lattice for an 
extended coupling scheme where the interaction decays exponentially.  The main graph is a semi-logarithmic plot of 
$(\delta_{\mathrm{RMS}}^{n})^{2}$ for various values of the decay constant $\gamma$, while the inset is the corresponding 
plot of $\delta_{\mathrm{rm}}^{n}$.}
\end{figure}

\begin{figure}
\includegraphics[width=.49\textwidth]{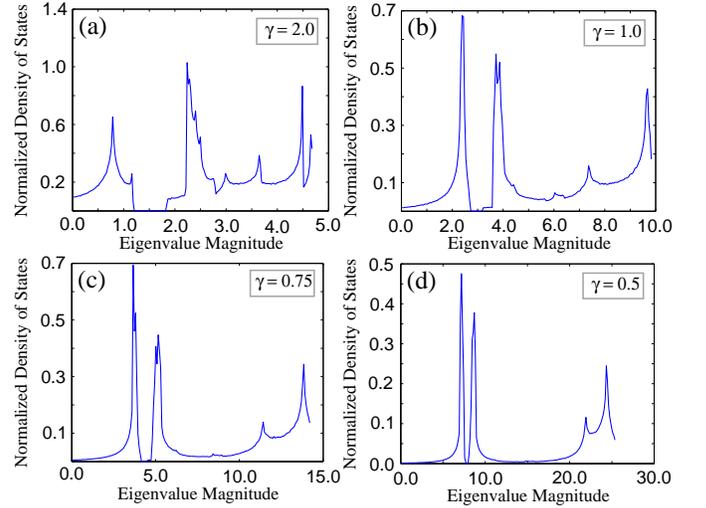}
\caption{\label{fig:Fig21} (Color Online) Density of states curves in the case of the 
honeycomb lattice for $\gamma = 2.0$,
$\gamma = 1.0$, $\gamma = 0.75$, and $\gamma = 0.50$ in panels (a), (b), (c), and (d) respectively.}
\end{figure}

\begin{figure}
\includegraphics[width=.49\textwidth]{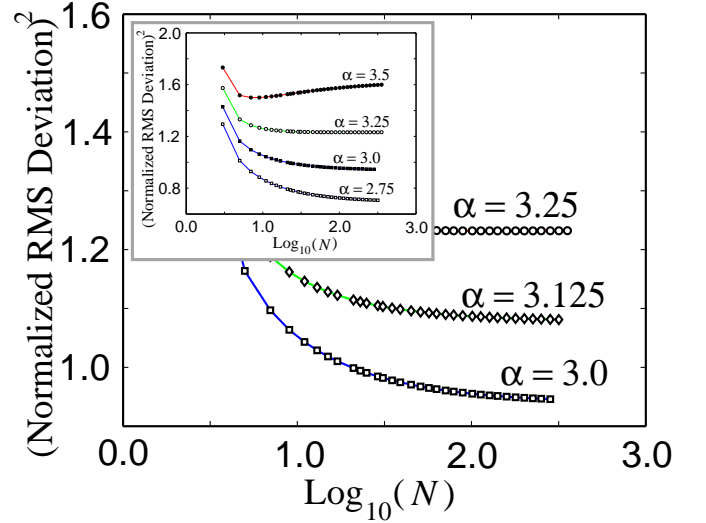}
\caption{\label{fig:Fig22} (Color Online) Mean square deviations for long-range interactions with a power law decay 
exponent $\alpha$ for the honeycomb lattice geometry. The inset shows a relatively broad range of $\alpha$ values, whereas the 
main graph is a closer view.}
\end{figure}

\begin{figure}
\includegraphics[width=.49\textwidth]{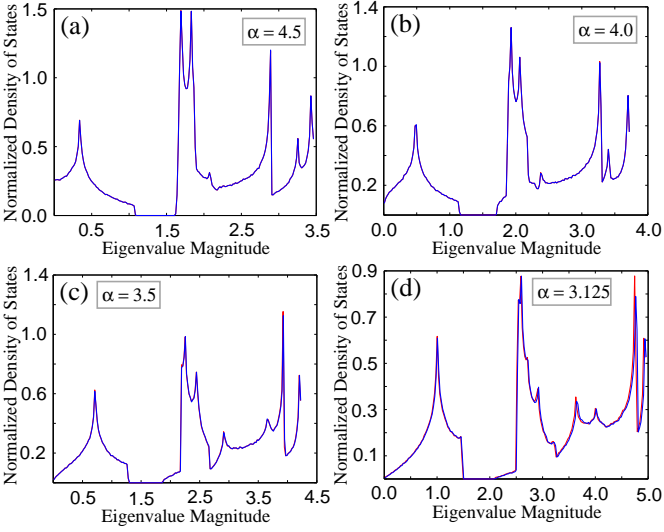}
\caption{\label{fig:Fig23} (Color Online) Density of states curves in the case of the 
honeycomb lattice with a power law coupling scheme for $\alpha = 4.5$,
$\alpha = 4.0$, $\alpha = 3.5$, and $\alpha = 3.125$ in panels (a), (b), (c), and (d) respectively.}
\end{figure}

\section{Transverse Displacements in an extended scheme}

In each of the preceding cases discussed in this work, thermally induced deviations of the    
lattice sites have been confined either to motion withing the lattice plane in the context of two 
dimensional systems, or collinear movements for the lattice in 1D.  However, we also consider 
displacements perpendicular to the lattice for two dimensional systems.   To provide local stiffness
of covalent two dimensional lattices realized in nature where the finite 
thickness would provide rigidity with respect to transverse displacements tending to push atoms above 
or below the plane of the crystal, we examine a dual layer geometry where an extended scheme 
provides a local stiffness. 

In fact, to provide as much stability as possible, we consider a long-range coupling scheme in 
the interaction between atoms within a layer as well as between layers decreases as a power law 
(with the decay exponent designated $\alpha$) in the 
separation between atomic species.

The potential energy stored in the lattice is similar in abstract form to the corresponding expression for the 
honeycomb lattices, and is given by
\begin{align}
E \! = \! \! \sum_{i,j = 0}^{N} \sum_{l,m=-n}^{n} \! \! \left( \! \! \! \begin{array}{c} \tfrac{1}{2} K(r_{lm}^{s})\! \left[ \!
\vec{\hat{\Delta}}_{lm}^{s}  \! \cdot \! (\vec{\delta}_{i+lj+m}^{A} \! - \!
\vec{\delta}_{ij}^{A} ) \! \right]^{2} \\ + \tfrac{1}{2} K(r_{lm}^{s}) \!
\left[ \! \vec{\hat{\Delta}}_{lm}^{s} \! \cdot \! (\vec{\delta}_{i+lj+m}^{B} \! - \!
\vec{\delta}_{ij}^{B} ) \! \right]^{2} \\ + \! \! K(r_{lm}^{\mathrm{ab}}) \left[ \! \vec{\hat{\Delta}}_{lm}^{\mathrm{ab}} 
 \! \cdot \! (\vec{\delta}_{i+lj+m}^{B} \! - \!
\vec{\delta}_{ij}^{A} ) \! \right]^{2} \end{array} \! \! \! \right)
\end{align}
in a real space representation, where the unit vectors corresponding the intraplanar couplings $\vec{\hat{\Delta}}_{lm}^{s}$ have 
$x$ and $y$ components given by 
\begin{align}
\hat{\Delta}_{lm}^{x(s)} = \frac{l}{r_{lm}^{s}};~~\hat{\Delta}_{lm}^{y(s)} = \frac{m}{r_{lm}^{s}},
\end{align}
where the separation between interacting sites within a plane is $r_{lm}^{s} = \sqrt{l^{2} + m^{2}}$
The components of the unit vector $\vec{\hat{\Delta}}_{lm}^{\mathrm{ab}}$ 
are identical to those of the planar case with the exception of a nonzero $z$ component $\hat{\Delta}_{lm}^{z(\mathrm{ab})} = 
1/r_{lm}^{\mathrm{ab}}$ where the distance between sites in the two distinct lattice planes is $r_{lm}^{\mathrm{ab}} = \sqrt{m^{2} + l^{2} + 1}$. 
 Operating in terms of Fourier components reduces the 
decoupling of the vibrational modes to the diagonalization of a $6 \times 6$ matrix of the form
where $\hat{A}$ and $\hat{B}$ are $3 \times 3$ sub-matrices, with 
$\hat{B}^{\dagger}$ the Hermitian conjugate of $\hat{B}$. The sub-matrices have the form      
\begin{align}
\hat{A} = \! \! \! \sum_{l,m=-n}^{n} \! \! \left[ \! \! \! \begin{array}{ccc} (d_{lm}^{xx(\mathrm{ab})} + d_{lm}^{xx(s)}) & (d_{lm}^{xy(\mathrm{ab})} + 
d_{lm}^{xy(s)}) & d_{lm}^{xz(\mathrm{ab})} \\
(d_{lm}^{yx(\mathrm{ab})} + d_{lm}^{yx(s)}) & (d_{lm}^{yy(\mathrm{ab})} + d_{lm}^{yy(s)}) & d_{lm}^{yz(\mathrm{ab})} \\
d_{lm}^{zx(\mathrm{ab}} & d_{lm}^{zy(\mathrm{ab}} & d_{lm}^{zz(\mathrm{ab})}
\end{array} \! \! \! \right ]
\end{align}
for $\hat{A}$, where for the sake of brevity we have used the notation, e.g.  $d_{lm}^{xy(\mathrm{ab})} \equiv K(r_{lm}^{\mathrm{ab}}) 
\hat{\Delta}_{lm}^{x(\mathrm{ab})} 
\hat{\Delta}_{lm}^{y(\mathrm{ab})}$ and $d_{lm}^{xy(s)} \equiv K(r_{lm}^{s}) (1 - \cos [k_{x} l + k_{y} m])
\hat{\Delta}_{lm}^{x(s)} \hat{\Delta}_{lm}^{y(s)}$.
The complex sub-matrix $\hat{B}$ is given by
\begin{align}
\hat{B} = \! \! \sum_{l,m=-n}^{n} e^{i(k_{x}l + k_{y} m)} 
\left[ \! \! \begin{array}{ccc} d_{lm}^{xx(\mathrm{ab})}  & d_{lm}^{xy(\mathrm{ab})} & d_{lm}^{xz(\mathrm{ab})} \\ 
d_{lm}^{yx(\mathrm{ab})}  & d_{lm}^{yy(\mathrm{ab})} & d_{lm}^{yz(\mathrm{ab})} \\
d_{lm}^{zx(\mathrm{ab})}  & d_{lm}^{zy(\mathrm{ab})} & d_{lm}^{zz(\mathrm{ab})} 
 \end{array} \! \! \right ]
\end{align}

\begin{figure}
\includegraphics[width=.49\textwidth]{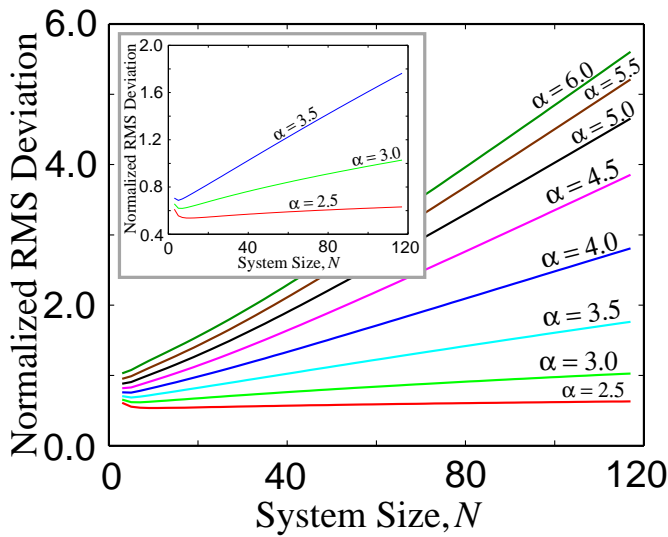}
\caption{\label{fig:Fig24} (Color Online) Mean square deviations for long-range interactions with a power law decay
exponent $\alpha$.}
\end{figure}

Results for the mean square deviation $\delta_{\mathrm{RMS}}^{n}$ are shown in Fig.~\ref{fig:Fig24} for a variety of 
decay exponents ranging from a weak decay $\alpha = -2.5$ to a considerably more rapid decay with 
the separation distance $\alpha = -6.0$.  A salient feature in each of the RMS curves 
shown in the graph is an asymptotic linear growth  
with the size $N$ of the system, though the slope of the linear diverge in system size decreases 
with decreasing $\alpha$.  Broadly speaking, there are two regimes for each value of $\alpha$ in the 
variation of the mean square deviation with system size.  For relatively small system sizes,  
$\delta_{\mathrm{RMS}}$ changes quite slowly with increasing $N$.  However, ultimately the RMS   
fluctuations begin to grow more rapidly and eventually   The size of the plateau where the mean square 
fluctuations expand slowly is slightly broader for small values of the decay exponent $\alpha$, and   
somewhat abbreviated for the more rapidly decaying coupling where $\alpha = -6$.  The latter is 
closer to what one would find in covalently bonded (but non-polar) systems such as graphene where 
Van der Waals interactions decreasing with the sixth power of the inter-atomic separation     
constitute the 
main source of long-range attraction between particles.  Hence, London interactions would not 
prevent atomic displacements transverse to the lattice plane from destroying 
crystalline order.

\section{Conclusions}
We have examined the effect of thermally induced lattice vibrations on long range order in
one and two dimensional crystals.
In the case of crystals in one dimension, long-range positional order is (as expected) disrupted by 
thermal fluctuations with the $\delta_{\mathrm{RMS}}^{n}$ scaling as $N^{1/2}$.  For inherently 
long-ranged interactions scaling as power laws in the distance between interacting atoms, 
the divergence is much slower for $\alpha > \alpha_{c}^{\mathrm{1D}} = 1.615$, while crystalline order is  
intact even at finite temperatures if $\alpha < \alpha_{c}^{\mathrm{1D}}$.

For two dimensional crystals, we find the same essential phenomena 
with respect to thermodynamic stability of the crystal for square, triangular, and 
honeycomb lattices.  For the latter two, thermal fluctuations destroy long-range crystalline 
order at finite temperatures, but the divergence in $\delta_{\mathrm{RMS}}^{n}$ occurs very 
slowly with increasing system size.  On the other hand, RMS deviations decay rapidly for 
simple square lattices where only nearest neighbor couplings are active.  However, an extended 
coupling scheme to both nearest and next-nearest neighbors considerably mitigates the effect 
of thermal fluctuations on crystalline order, and the resulting slow divergence is    
quantitatively similar to that seen in the triangular and honeycomb lattices where the coupling  
is confined entirely to nearest-neighbors. 

When we extend the coupling to many neighbors, but still implement a short-ranged coupling 
(e.g. an exponential decay with the range set by the inverse decay rate $\gamma^{-1}$), 
we find qualitatively the same results as for the square lattice with couplings both to 
nearest and next-nearest neighbors, as well as the triangular and hexagonal lattices in 
atomic members only interact with nearest neighbors with
$(\delta_{\mathrm{RMS}}^{n})^{2}$ scaling linearly with $\log_{10}(N)$.  However, the 
slope of the linear dependence  
becomes smaller as $\gamma$ is decreased, as a longer-ranged coupling is more effective at 
suppressing the effects of thermal fluctuations.
As in the case of systems in 1D, a longer range coupling in the form of a power law 
can maintain long-range crystalline order for $T > 0$ if the decay exponent does not 
exceed $\alpha_{c}^{\mathrm{2D}} = 3.15$.

Allowing thermally induced fluctuations perpendicular to the lattice causes a rapid divergence 
(i.e. linear) of the RMS deviations with the system size, even if the lattice geometry is dual-layered 
in an extended scheme to provide local stiffness.  
The growth of $\delta_{\mathrm{rms}}^{n}$ is asymptotically linear even if the coupling between sites is 
long-ranged, decaying as a power law.  Hence, long-ranged London interactions would not be enough by themselves
to preserve positional order in a covalently bonded locally rigid two dimensional lattice.

\begin{acknowledgments}
Useful conversations with Yogesh Joglekar are gratefully acknowledged.
\end{acknowledgments}


\end{document}